\documentclass[11pt,a4paper]{article}
\usepackage{jheppub}
\usepackage{amsmath,amsfonts,amssymb,amsthm,graphics,graphicx,epsfig,times,bbm,verbatim}

\usepackage{rotating}
\usepackage{rotfloat}
\usepackage{mathrsfs} 
\usepackage{booktabs} 

\def\beqn{\begin{eqnarray}}
\def\eeqn{\end{eqnarray}}
\def\beqs{\begin{subequations}}
\def\eeqs{\end{subequations}}

\newcommand\be{\begin{equation} }
\newcommand\ee{\end{equation}}
\newcommand\IC{{\mathbb{C}}}
\newcommand\IH{{\mathbb{H}}}
\newcommand\II{{\mathbb{I}}}
\newcommand\IK{{\mathbb{K}}}
\newcommand\IQ{{\mathbb{Q}}}

\newcommand\IZ{{\mathbb{Z}}}

\newcommand{\me}{\mathrm{e}}
\newcommand{\mi}{\mathrm{i}}
\newcommand{\md}{\mathrm{d}}

\def\CA{{\cal A}}\def\CD{{\cal D}}
\def\CF{{\cal F}}\def\CG{{\cal G}}
\def\CH{{\cal H}}\def\CI{{\cal I}}
\def\CM{{\cal M}}
\def\CN{{\cal N}}\def\CP{{\cal P}}

\def\CW{{\cal W}}
\def\CC{{\cal C}}

\title{Galois Symmetry Induced by Hecke Relations in Rational Conformal Field Theory and Associated Modular Tensor Categories}

\author[1]{Jeffrey A. Harvey,}
\author[2]{Yichen Hu}
\author[1]{and Yuxiao Wu}
\affiliation[1]{Enrico Fermi Institute and Department of Physics, University of Chicago, \\
933 E 56th St., Chicago, IL 60637}
\affiliation[2]{Rudolf Peierls Centre for Theoretical Physics, Clarendon Laboratory, \\
Parks Road, Oxford, OX1 3PU, United Kingdom}
\emailAdd{j-harvey@uchicago.edu}
\emailAdd{yichen.hu@physics.ox.ac.uk}
\emailAdd{yuxiaowu@uchicago.edu}

\abstract{Hecke operators relate characters of rational conformal field theories (RCFTs) with different central charges, and extend the previously studied Galois symmetry of modular representations and fusion algebras. 
We show that the conductor $N$ of a RCFT and the quadratic residues modulo $N$ play an important role in the computation and classification of Galois permutations.
We establish a field correspondence in different theories through the picture of effective central charge, which combines Galois inner automorphisms and the structure of simple currents.
We then make a first attempt to extend Hecke operators to the full data of modular tensor categories.
The Galois symmetry encountered in the modular data transforms the fusion and the braiding matrices as well, and yields isomorphic structures in theories related by Hecke operators.
\vskip 0.1in
\today}


\begin{document}

\maketitle

\section{Introduction}
\label{sec:Intro}

Two-dimensional rational conformal field theories (RCFTs) have found applications in the worldsheet description of classical string theory backgrounds, as well as in many areas in condensed matter physics such as quantum Hall systems and the study of  boundary modes in topological insulators.
The characters of RCFT are partition functions on the torus, and record the number of physical states.
Because of the modular properties under the action of $SL(2,\IZ)$, the characters are also modular functions and thus also encode fascinating number theoretic features.

Recently it has been discovered that Hecke operators relate characters of  certain RCFTs with different central charges \cite{Hecke:2018}.
These Hecke operators extend the known Galois symmetry connecting modular representations.
They act on vector-valued modular functions which may be characters of one RCFT and often produce characters of another RCFT
which is not obviously related to the original RCFT. Two RCFTs whose characters are related by Hecke operators are clearly not the same RCFT since they have different central charges,  but it could be that some algebraic structure related to the two RCFTs is the same. In particular we will provide evidence that the modular tensor categories (MTCs) related to the two RCFTs are either the same or closely related.\footnote{We thank S. Gukov and G. Moore for suggesting that we investigate the relation between Hecke operators and MTCs.} We explore this possibility in a number of simple cases in this paper.

MTCs arise as representation categories and encode the topological structures of vertex operator algebras (VOAs) 
and CFTs \cite{Moore:1989vd,Moore:polynomial,Moore:classical,Turaev:book}.
MTCs of low rank are classified in \cite{Gepner:fusion_ring,Rowell:2009}. See \cite{VOA_database} for a catalog of known MTCs. 
The basic data in a MTC include the twists (topological spins) which are exponentials of the conformal weights
\begin{align}
\theta_i=\me(h_i),
\end{align}
and the quantum dimensions $d_i$ which are  ratios of elements of the modular $S$ matrix
\begin{align}\label{def:q_dim}
d_i=\rho(S)_{0i}\big/\rho(S)_{00}.
\end{align}
Here we are using the standard convention in number theory that $\me(x) \equiv \me^{ 2 \pi \mi x}$. 
When the MTC is unitary, $\rho(S)_{00}$ is positive and the $d_i$ are positive numbers greater than or equal to 1.
The (topological) central charge $c$  is related to the twists and the quantum dimensions by
\begin{align}
\me\Big(\frac{c}{8}\Big) = 
\rho(S)_{00} \sum_i \theta_i d_i^2.
\end{align}
A MTC may arise from more than one RCFT, since the MTC only fixes $h_i \pmod{1}$ and $c \pmod{8}$. 
A MTC is also equipped with  duality matrices obeying the consistency conditions known as pentagon and the hexagon identities \cite{Moore:polynomial, Moore:classical}. 
Extensive applications of MTC are found in condensed matter physics, where they offer tools for studying anyonic systems and topological quantum computation \cite{Moore_Read:1991,Wen:2015,Schoutens:2016,Bonderson:thesis}.

Our first main result concerns the structure of the Galois permutations induced by Hecke operators. All the Galois information is traced back to the conductor $N$, and the unit group $(\mathbb{Z}/N\mathbb{Z})^{\times}$ can be represented by the Frobenius maps.
Upon the Hecke operation, a Frobenius map acts on the modular representation. However the permutations are characterized by the quadratic residues in $(\mathbb{Z}/N\mathbb{Z})^{\times}$, in other words the quadratic subgroup determines the Galois group of fusion rules.
With effective central charges less than 1, the Virasoro minimal models are nice candidates to probe the Hecke images of the characters, as well as Galois conjugates of modular representations. A number of examples are presented in Section \ref{sec:Hecke_review}.  We note that the modular representation of the minimal model $\mathsf{M}(2,k+2)$ coincides with the $(-k)$-th Galois conjugate of $SU(2)_k$. Consequently, $\mathsf{M}(2,k+2)$ has identical fusion rules as  $SU(2)_k$ when restricted to integer spins, thus explaining this observation in condensed matter physics.

The second main result is that the RCFTs related by Hecke operators embody Galois symmetry in their fusion and braiding matrices. 
Given a RCFT and its MTC, the Hecke image of the characters gives rise to a Galois conjugate of the initial MTC. As the structure of MTC contains the duality transformation of conformal blocks, the Galois symmetry applies to the duality property naturally.  
To interpret this phenomenon, we exploit the picture of effective central charge as an intermediate step, and show that the initial RCFT and the Hecke image theory share identical fusion rules.
The fusion and braiding matrices in each image theory obey the same pentagon and hexagon system of equations, whose different solutions are related by Galois symmetry.

Recently a number of papers have appeared concerning the action of Hecke operators on vector-valued modular forms, see \cite{mwr,stein, diacon,Bouchard:Hecke} for details. These results are related to ours, but none seems to coincide precisely with the Hecke operators
defined in \cite{Hecke:2018}. 

This paper is organized as follows.
In Section \ref{sec:Hecke_review} we review the general structure of RCFT, give the definition of Hecke operators for $\Gamma(N)$, and discuss Galois permutations.
We then describe in detail the Hecke images and the Galois symmetry in several examples of RCFT and MTC.
Section \ref{sec:eff_picture} introduces the picture of effective central charge, which facilitates the derivation of the fusion rules of the Hecke image theory.
In Section \ref{sec:duality} we review duality transformations in RCFT, which include both fusing and braiding. 
We then turn in Section \ref{sec:Fusion_and_Braiding} to the Galois symmetry on fusion and braiding matrices of the Hecke image theory.
Finally in Section \ref{sec:Conclusions}, we conclude and suggest relevant problems for future study.

\section{Hecke Operators and Galois Symmetry}
\label{sec:Hecke_review}

We first give a brief introduction of RCFT characters and modular symmetry before defining  Hecke operators.
We refer the reader to \cite{Moore:1989vd, Fuchs:2009iz, Huang:2013jza} for an overview of RCFT.

In a two-dimensional RCFT, the Hilbert space decomposes into a finite sum of irreducible representations $V_i, \overline{V_{\bar i}}$ of the chiral algebras ${\cal A}$ and $\overline{\cal A}$, namely
\begin{align}\label{H:decomp}
{\cal H} = \bigoplus_{i \in {\cal I}, \bar i \in \bar {\cal I}} {\cal N}_{i, \bar i} \, V_i \otimes \overline{V_{\bar i}} ~,
\end{align}
where ${\cal N}_{i \bar i} \in \IZ_{\ge 0}$ and ${\cal I}$, $\bar {\cal I}$ are finite index sets labelling irreducible representations of ${\cal A}$ and $\overline{\cal A}$. In each representation $V_i$, one has the character
\begin{align}
\chi_i(\tau) = {\rm Tr}_{V_i} q^{L_0-c/24}, \qquad q=\me(\tau) ,\quad \tau \in \IH,
\end{align}
where $c$ is the central charge and $\IH$ is the upper half complex plane.
Following the decomposition of $\CH$, the partition function is a sesquilinear form of the characters $\chi_i(\tau)$:
\begin{align}
Z = \sum_{i \in {\cal I},  \bar i \in \bar {\cal I}} {\cal N}_{i, \bar i} \, \chi_i(\tau) \overline{ \chi_{\bar i}(\tau) } \, .
\end{align}

The full modular group $SL(2,\IZ)$ acts on $\tau$ in the upper half plane $\IH$ by
\be
\tau \rightarrow \gamma \tau := \frac{a \tau +b}{c \tau +d}
\ee
with
\be
\gamma = \begin{pmatrix} a & b \\ c & d \end{pmatrix} \in SL(2,\IZ) \, .
\ee
We call $f(\tau)$ a  (weakly holomorphic) modular form of weight $k$ for the modular group $\Gamma=SL(2,\IZ)$ if $f$: $\IH \rightarrow \IC$ is holomorphic (except for a possible pole as $\tau \rightarrow \mi \infty$) in $\IH$ and obeys the transformation law
\be \label{ftransf}
f( \gamma \tau) = (c \tau+d)^k f(\tau) \, .
\ee
It suffices to know the action by the $SL(2,\IZ)$ generators
\be
T= \begin{pmatrix} 1 & 1 \\ 0 & 1 \end{pmatrix}, \qquad S= \begin{pmatrix} 0 & -1 \\ 1 & 0 \end{pmatrix}.
\ee
In RCFT, the individual characters $\chi_i(\tau)$ are weakly holomorphic modular functions for the principal congruence subgroup $\Gamma(N)$ for a finite $N$ defined below. Under the $SL(2,\IZ)$ transformation $\gamma$, the characters transform as
\begin{align}
\chi_i(\gamma \tau) = \sum_j \rho(\gamma)_{i j} \,\chi_j(\tau) \, .
\end{align}
Here, $\rho$ is a finite-dimensional representation of $SL(2,\IZ)$:
\begin{align}
\rho: SL(2, \IZ) \rightarrow GL(V),
\end{align}
which is completely determined by its values on the $SL(2,\IZ)$ generators 
\begin{align}
S: \tau \rightarrow -1/\tau ,\qquad
T: \tau \rightarrow \tau+1.
\end{align}
The partition function $Z$ must be modular invariant.
As an $SL(2, \IZ)$ representation, $\rho$ obeys the consistency condition
\begin{align}
\rho(S)^2 =
\big( \rho(T)\rho(S) \big)^3 =\CC ,
\end{align}
where $\CC$ is the charge conjugation matrix.
In \cite{Hecke:2018} the first and third authors studied the Hecke operators for $\Gamma(N)$ modular forms.
These Hecke operators act nicely on the Fourier expansion of the characters
\begin{align}
\chi_i(\tau)=q^{h_i-\frac{c}{24}}\sum_{n=0}^{\infty} a_i(n) \, q^n,
\qquad q=\me(\tau),
\end{align}
where $c$ is the central charge and $h_i$ is the conformal weight.

The fusion coefficients ${}_0{N_{ij}}^k$ which govern the fusion of primary operators $\phi_i$ as
\be\label{fusion_rule}
\phi_i \times \phi_j = \sum_k {}_0{N_{ij}}^k \, \phi_k \, 
\ee
are determined by the Verlinde formula
\begin{align}\label{Verlinde_formula}
{}_0{N_{ij}}^k = \sum_m \frac{\rho(S)_{im} \rho(S)_{jm} \rho(S^{-1})_{km}}{\rho(S)_{0m}} 
\end{align}
\cite{Verlinde:1988sn}, where the label $0$ emphasizes the special role played by the vacuum entry \cite{Fuchs:1994}. The fusion coefficients in a unitary RCFT must be non-negative.
The fusion rules for the field $i$ are gathered into the matrix $N_i$ with the element
\begin{align}\label{fusion_rule:matrix}
(N_i)_{j,k} = {}_0{N_{ij}}^k .
\end{align}

\subsection{Hecke operators for $\Gamma(N)$}
\label{subsec:Hecke_Gamma(N)}

Since $\chi_i(\tau)$ is a $q$-series with leading term $q^{h_i- c/24}$, the matrix $\rho(T)$ is diagonal with entries $\me(h_i- c/24)$.
In RCFT the conformal weights $h_i$ and the central charge $c$ are rational \cite{Anderson:1987ge}. Hence $\rho(T)$ has a finite order $N$, which is the least common denominator of $h_i- c/24$.
Theorem 1 of Bantay \cite{Bantay:2001ni} states that the kernel of $\rho$ contains the principal congruence subgroup $\Gamma(N)$ defined as
\begin{align}
\Gamma(N) =  \left \{ \begin{pmatrix} a & b \\ c & d \end{pmatrix} \in SL(2,\IZ) \Bigg| ~ a \equiv d \equiv 1 \pmod{N}, ~b \equiv c \equiv 0 \pmod{N} \right \} \, .
\end{align}
In other words, $\chi_i(\tau)$ are invariant under $\tau \rightarrow \gamma \tau$ for $\gamma \in \Gamma(N)$.
There is a natural homomorphism 
\begin{align}
\mu_N: SL(2,\mathbb{Z}) \rightarrow SL(2,\mathbb{Z}/N\mathbb{Z})
\end{align}
done by reduction mod $N$ of each element $\gamma \in SL(2,\IZ)$. Because the kernel of $\mu_N$ is precisely $\Gamma(N)$, the map $\mu_N$ does not affect the modular representation. Hence, $\rho$ can be also regarded as a representation of $SL(2,\mathbb{Z}/N\mathbb{Z})$.

The Hecke operator $T_p$ for $SL(2,\mathbb{Z})$ modular forms has been discussed in textbooks on number theory \cite{serre, zagier}. However, characters in RCFT are modular functions for $\Gamma(N)$ and transform according to the representation $\rho$ under $SL(2, \mathbb{Z})$, that is they are vector-valued forms rather than strictly modular forms. The Hecke operators on them should be compatible with their vector structure. 
To define the Hecke operators for $\Gamma(N)$, we introduce the set of orbit representatives 
\begin{align}\begin{split}
\label{rep_gen}
\Delta_N^{(p)} &=
\left\{
\sigma_p
\begin{pmatrix}
p & 0 \\ 0 & 1
\end{pmatrix}, 
\begin{pmatrix} 1 & bN \\0 & p \end{pmatrix}
\Big|~
0\leq b \leq p-1
\right\} \\
&= \sigma_{p} \circ
\left\{
\begin{pmatrix}
p & 0 \\ 0 & 1
\end{pmatrix}, 
\sigma_{\bar p} 
\begin{pmatrix} 1 & bN \\0 & p \end{pmatrix}
\Big|~
0\leq b \leq p-1
\right\},
\end{split}\end{align}
where $p$ is a prime number with $\text{gcd}(p,N)=1$ \cite{Rankin}. Here $\sigma_p$ denotes the preimage of
\begin{align}
\begin{pmatrix} 
\bar p & 0 \\ 0 & p 
\end{pmatrix}
\end{align}
under $\mu_N$, and $\bar p$ is the multiplicative inverse of $p$ modulo $N$. The occurrence of $\sigma_p$ reflects the nature of $\Gamma(N)$. 
Properties of $\rho(\sigma_p)$ will be addressed shortly.

Define the Hecke operator $\mathsf{T}_p$ acting on weight zero vector-valued modular form $f$ relative to a representation $\rho$ for $p$ prime 
\begin{align} \label{N_Hecke:define}
(\mathsf{T}_p f)_i(\tau) :=  \sum_{ \delta \in \Delta_N^{(p)}} f_i(\delta \tau) = \sum_j \rho_{ij}(\sigma_p) f_j(p \tau) +\sum_{b=0}^{p-1} f_i\left(\frac{\tau+bN}{p}\right) 
\end{align}
\cite{Hecke:2018}. The normalization of $\mathsf{T}_p$ differs from the traditional $T_p$ for scalar modular forms in order to preserve the integrality of coefficients.
Given the form of $\mathsf{T}_p$ for $p$ prime, one can construct Hecke operators $\mathsf{T}_n$ for $n$ coprime to $N$ but not necessarily prime.
See the appendix of \cite{Hecke:2018} for more detail.

An essential ingredient in defining Hecke operators for $\Gamma(N)$ is the representation matrix of the $SL(2,\IZ)$ element $\sigma_p$, which also constitutes the modular representation of the Hecke image.
Under the Hecke operation $\mathsf{T}_p$, the induced modular representation $\rho^{(p)}$ is related to the original representation $\rho$ via 
\begin{align} \label{newrep}
\rho^{(p)}(T)  = \rho(T^{\bar p}), \qquad \rho^{(p)}(S)= \rho( \sigma_p S)  .
\end{align}
Though $\sigma_p$ is not unique, two choices differ by the action of $\Gamma(N)$ which is in the kernel of $\rho$. For this reason, the representation $\rho^{(p)}$ is uniquely determined by any choice of $\sigma_p$.
Since $\text{gcd}(\bar p,N)=1$, $\rho^{(p)}(T)$ has the same order as $\rho(T)$. Therefore the action of $\mathsf{T}_p$ preserves the value of $N$ and the dimension of the representation.

\subsection{Galois permutations}
\label{subsec:Galois_Permu}
A crucial consequence of the Hecke operation is the induced Galois symmetry, which relates $SL(2, \IZ)$ representations and thereby fusion rules in RCFTs. In a nondegenerate RCFT (the conformal weights of primary fields do not differ by integers), we show that the Galois group of fusion rules is fully determined by the conductor. 

Denote by $\mathbb{K}$ the number field obtained by adjoining all matrix elements of the modular representation to $\IQ$. De Boer and Goeree show that $\mathbb{K}$ is a finite Abelian extension of $\IQ$ \cite{DeBoer:1990em}.
Write $\xi_m=\me(1/m)$. Denoting by $\IQ[\xi_m]$ the cyclotomic field that is an extension of $\IQ$ by a primitive $m$th root of unity, the smallest integer $m$ such that $\IK \subset \IQ[\xi_m]$ is called the conductor of the RCFT. 
Bantay shows that the conductor equals precisely $N$, the order of $\rho(T)$ \cite{Bantay:2001ni}.
Moreover, since $\IK$ contains the $N$th roots of unity as the diagonal entries of $\rho(T)$, $\IK$ is exactly $\IQ[\xi_N]$.
The automorphisms of $\IK$ over $\IQ$ furnish the Galois group $\CG_N=\text{Gal}(\mathbb{Q}[\xi_N]/\mathbb{Q})$, which is isomorphic to the unit group $(\mathbb{Z}/N\mathbb{Z})^{\times}$, the group of multiplicative units in $\mathbb{Z}/N\mathbb{Z}$. Each element $\ell$ in $(\mathbb{Z}/N\mathbb{Z})^{\times}$ gives rise to a Frobenius map $f_{N,\ell}$ which takes $\xi_N$ to $\xi_N^{\ell}$ while leaving $\mathbb{Q}$ fixed. 

We write $\bar p$ for the multiplicative inverse of $p$ in $(\mathbb{Z}/N\mathbb{Z})^{\times}$.
As discussed in \cite{Coste:1993af}, the Frobenius element $f_{N,\bar p}$ acts on the representation matrices $\rho(T),\rho(S)$ as
\beqs
\begin{align}
f_{N,\bar p} \big(\rho(T) \big)&= \rho(T)^{\bar p}, \\
f_{N,\bar p} \big(\rho(S) \big)&= \rho(S) \, G_p=G_p^{-1} \, \rho(S) .
\end{align}
\eeqs
The matrix $G_p$ coincides with $\rho(\sigma_{\bar p})$ \cite{Hecke:2018},
proving that the modular representation $\rho^{(p)}$ is equivalent to the Galois action $f_{N,\bar p}$ on $\rho$:
\begin{align} \label{rep:Frobenius}
f_{N,\bar p}\big(\rho(T)\big) 
=\rho^{(p)}(T), 
\qquad f_{N,\bar p}\big(\rho(S)\big) = \rho^{(p)}(S) .
\end{align}
The Hecke operator $\mathsf{T}_p$ extends $\rho^{(\bar p)}$ to an action on the characters of the RCFT rather than just on the modular representation.

Let $N=\prod_{i=1}^r p_i^{k_i}$ be the prime factorization of the conductor.
The matrices $G_p=\rho(\sigma_{\bar p})$ reveal intriguing features as the representation of $SL(2,\IZ/N\IZ)$.
The finite-dimensional representation $\rho$ is completely reducible, and each irreducible component $\omega$ of $\rho$ has the unique product decomposition 
\begin{align}
\omega = \otimes_{i=1}^n \pi\big(p_i^{k_i}\big)
\end{align}
\cite{eholzerII}. Here $\pi\big(p_i^{k_i}\big)$ is an irreducible representation of $SL \big(2,\mathbb{Z}/p_i^{k_i}\mathbb{Z} \big)$.
The homomorphism $p \rightarrow \rho(\sigma_p)$ defines an $n$-dimensional representation  of $(\mathbb{Z}/N\mathbb{Z})^{\times}$, where $n=|\CI|$ and $\CI$ is the finite index set which labels the irreducible representations of the chiral algebra.

An explicit computation gives
\begin{align}
 T^p\, S^{-1}\, T^{\bar p} \, S \, T^p \, S  =
\left(
\begin{array}{cc}
 \left(1-p \bar{p}\right) p+p & p \bar{p}-1 \\
 1-p \bar{p} & \bar{p} \\
\end{array}
\right)
\equiv 
\begin{pmatrix}
p & 0 \\ 0 & \bar p
\end{pmatrix}
\pmod{N} ,
\end{align}
which establishes that $T^p\, S^{-1}\, T^{\bar p} \, S \, T^p \, S$ is the preimage of $\sigma_{\bar p}$ under the mod $N$ map from $SL(2,\IZ)$ to $SL(2,\IZ/N \IZ)$. In practice, $G_p$ can be evaluated from the expression
\begin{align}\label{Gp:formula}
G_p =\rho \big( T^p\, S^{-1}\, T^{\bar p} \, S \, T^p \, S \big)
=  \rho(T)^p\, \rho (S)^{-1}\, \rho(T)^{\bar p} \, \rho (S) \, \rho(T)^p \, \rho (S)  \, .
\end{align}
As it turns out, $G_p$ is a monomial matrix with the elements
\begin{align}\label{G_permu}
(G_p)_{i,j} =\varepsilon_p(i)\,\delta_{\pi_p (i),j}~,
\end{align}
where $\pi_p$ is some permutation of ${\cal I}$ and $\varepsilon_p$ is a map from ${\cal I}$ to $\{+1, -1 \}$ \cite{Coste:1999yc,Coste:1993af}.
When $p\equiv l^2 \pmod{N}$ for some $l \in (\mathbb{Z}/N\mathbb{Z})^{\times}$, $f_{N,p}$ induces the inner automorphism of $\rho(\gamma)$:
\begin{align}\label{inner_auto}
f_{N,p}\big( \rho(T) \big)= G_l \, \rho(T) \, G_l^{-1},\qquad
f_{N,p}\big( \rho(S) \big)= G_l \, \rho(S) \, G_l^{-1}~.
\end{align}
Hence, $f_{N,l^2}$ shuffles the diagonal entries of $\rho(T)$ by
\begin{align}\label{T_l^2 entry}
\rho(T)^{l^2}_{aa}=\rho(T)_{\pi_l(a);\pi_l(a)} ~, \qquad a \in \CI.
\end{align}
The permutations $\pi_p$'s encoded in $G_p$'s form the Galois group of fusion rules, denoted by $\CG$.
By the Galois group of fusion rules, we mean the automorphisms on the fusion rules which are caused by similarity transformations on the modular representation. We will see shortly that this Galois group consists of quadratic elements in $(\mathbb{Z}/N\mathbb{Z})^{\times}$ and is a subgroup thereof.
It should not be confused with the larger Galois group $\CG_N$ which acts on the cyclotomic number field $\mathbb{Q}[\xi_N]$ and is isomorphic to $(\mathbb{Z}/N\mathbb{Z})^{\times}$.

Next we demonstrate that the Galois permutations are determined by the quadratic residues modulo $N$.
For $l_1^2\equiv l_2^2 \pmod{N}$, one verifies that
\begin{align}
G_{l_1} \,\rho(T)\, G_{l_1}^{-1}
=G_{l_2} \,\rho(T)\, G_{l_2}^{-1},
\end{align}
and finds that $\rho(T)$ is invariant under conjugation by $G_{\bar l_1 l_2}$, i.e.
\begin{align}\label{permu1:bar l1 l2}
\rho(T) = G_{\bar l_1 l_2} \,\rho(T)\, G_{l_1 \bar l_2} \, .
\end{align}
In terms of the permutation, this equation states that
\begin{align}\label{permu2:bar l1 l2}
\rho(T)_{aa} = \rho(T)_{\pi_{\bar l_1 l_2}(a);\pi_{\bar l_1 l_2}(a)}.
\end{align}
The permutation $\pi_{\bar l_1 l_2}$ only shuffles the fields with same twist.
In most cases we encounter nondegenerate modular fusion algebras where independent characters are associated with different twists $\theta_i$, and thus all the eigenvalues of $\rho(T)$ are distinct \cite{eholzerII}.
As a result, $\pi_{\bar l_1 l_2}$ is an identity permutation, and $G_{\bar l_1 l_2}$ must be diagonal with entries $\pm 1$. 
Lemma 5 in \cite{Bantay:2001ni} states that if $G_p$ is diagonal then it must be $\pm \II$, hence we deduce
\begin{align}
G_{\bar l_1 l_2} = \pm \II  ,
\qquad \text{for } (\bar l_1 l_2)^2 \equiv 1 \pmod{N}.
\end{align}
This reasoning leads to the following result:

{\it 
In a nondegenerate modular fusion algebra we have the relation
\begin{align}
G_{l_1}=\pm G_{l_2}, 
\end{align}
if $l_1^2\equiv l_2^2 \pmod{N}$. In particular, $G_l=\pm \mathbb{I}$ for every $l$ such that $l^2\equiv 1 \pmod{N}$.
}
A modular fusion algebra is called nondegenerate when the conformal weights of a possible underlying RCFT do not differ by integers \cite{eholzerII}.

This shows that $G_l$ is specified by the congruence class of $l^2$ modulo $N$, up to a parity sign $\varepsilon_l(0)$ which does not affect the Galois permutation $\pi_l$.
In some RCFTs there can be complex-conjugate primary fields which have the same character.
We may regard them as a single neutral primary and reduce the dimensionality of the modular representation, prior to imposing the nondegeneracy condition. Then the Hecke operators will act on the reduced vector-valued modular form. See examples in Section \ref{subsec:higher_rank}.

In applying the above result it is useful to discuss the structure of $(\mathbb{Z}/N\mathbb{Z})^{\times}$ and the group of quadratic residues modulo $N$.
With the prime factorization $N=\prod_{i=1}^r p_i^{k_i}$, the unit group $(\mathbb{Z}/N\mathbb{Z})^{\times}$ is the direct product of the unit groups associated with each prime power factor
\begin{align}\label{factorization_1}
(\mathbb{Z}/N\mathbb{Z})^{\times} \cong \prod_{i=1}^r \big( \mathbb{Z}/p_i^{k_i}\mathbb{Z} \big)^{\times} 
\end{align}
by the Chinese Remainder Theorem. 
Each prime sector can be expressed by the cyclic group $C_m$.
\begin{align}
\big( \mathbb{Z}/p^k \mathbb{Z} \big)^{\times} \cong
\begin{cases} 
1
& \text{if  }  p=2 \text{ and } k=1 , \\
C_2 \times C_{2^{k-2}}
&  \text{if  }  p=2 \text{ and } k \geq 2 , \\
C_{p^{k-1}(p-1)}
&  \text{if  }  p>2  .
\end{cases}  
\end{align}
Define the group of quadratic residues modulo $N$
\begin{align}\label{quadratic_residue}
\big[(\mathbb{Z}/N\mathbb{Z})^{\times}\big]^2
:=\big \{m^2~|~ m\in (\mathbb{Z}/N\mathbb{Z})^{\times} \big\},
\end{align}
which is evidently a subgroup of $(\mathbb{Z}/N\mathbb{Z})^{\times}$. The group
$\big[(\mathbb{Z}/N\mathbb{Z})^{\times}\big]^2$ can be calculated by folding the components in eq\eqref{factorization_1}.

A wide class of RCFTs are the Virasoro minimal models $\mathsf{M}(p_1,p_2)$, which are labeled by a pair of coprime integers $(p_1,p_2)$ with $p_1,p_2>1$. The model $\mathsf{M}(p_1,p_2)$ is unitary iff $|p_1 - p_2|=1$.
Both unitary and non-unitary minimal models will be considered in this work.
Their conductors are computed in \cite{Bantay:2001ni}.
We list in Table \ref{table:data} the unit group $(\mathbb{Z}/N\mathbb{Z})^{\times}$ and the group of quadratic residues $\big[(\mathbb{Z}/N\mathbb{Z})^{\times}\big]^2$ for a number of minimal models, as well as their Galois groups.

Affine Lie algebras (Kac-Moody algebras) are also important examples of RCFT.
They are infinite dimensional algebras that extend simple Lie algebras, and appear as current algebras in the WZW models. 
In an affine Lie algebra $(G,k)$, the integer $k$ denotes the central extension called the level.
The Virasoro algebra is supplemented by the holomorphic spin-1 currents that satisfy the commutation relations
\begin{align}
\big[ J^a_m,J^b_n \big]= \mi \sum_c f_{abc} \, J^c_{m+n} 
+ k\,m\, \delta^{ab}\delta_{m+n,0}~,
\end{align}
where $f_{abc}$ are the structure constants of the Lie algebra $G$.
The characters of an affine Lie algebra transform among themselves under the modular group. In affine Lie algebras there is a Galois symmetry acting on highest weight representations \cite{Fuchs:1994,Fuchs:1995}, and the resulting fusion rule automorphism is discussed in  \cite{Coste:1993af}.

\begin{table}
\centering
\begin{tabular}{|c|c|c|c|c|c|}
\hline \hline
$(p_1,p_2)$  & $N$  & $n$ & $(\mathbb{Z}/N\mathbb{Z})^{\times}$  & $\big[(\mathbb{Z}/N\mathbb{Z})^{\times}\big]^2 $ & $\CG$ \\ 
\hline \hline
$(2,5)$  &  $60$   &  $2$ &  $C_2\times C_2 \times C_4$  & $C_2$ & $C_2$\\ 
\hline
$(2,7)$  &  $42$   &  $3$ &  $C_2 \times C_6$  & $C_3$ & $C_3$\\ 
\hline
$(2,9)$  &  $36$   &  $4$ &  $C_2 \times C_6$  & $C_3$ & $C_3$\\ 
\hline
$(2,11)$  &  $33$   &  $5$ &  $C_2 \times C_{10}$  & $C_5$ & $C_5$\\ 
\hline
$(2,13)$  &  $156$   &  $6$ &  $C_2 \times C_2 \times C_{12}$  & $C_6$ & $C_6$\\ 
\hline
$(2,15)$*  &  $30$   &  $7$ &  $C_2 \times C_4$  & $C_2$ & $C_4$ \\ 
\hline
$(2,19)$  &  $57$   &  $9$ &  $C_2 \times C_{18}$  & $C_9$ & $C_9$\\ 
\hline
$(3,4)$  &  $48$   &  $3$ &  $C_2 \times C_4 \times C_2$  & $C_2$ & $C_2$ \\ 
\hline
$(3,5)$  &  $40$   &  $4$ &  $C_2 \times C_2 \times C_4$  & $C_2$ & $C_2$ \\ 
\hline
$(3,7)$  &  $168$   &  $6$ &  $C_2 \times C_2 \times C_2 \times C_6$  & $C_3$ & $C_3$ \\ 
\hline
$(3,8)$  &  $32$   &  $7$ &  $C_2 \times C_8$  & $C_4$ & $C_4$\\ 
\hline\hline
\end{tabular}
\caption{\label{table:data} This table summarizes some data for a few minimal models. $N$ is the conductor of the theory, which is also equal to the order of $\rho(T)$. $n=| \CI |$ denotes the number of primary fields. $\CG$ stands for the Galois group of fusion rules.
We put an asterisk on $\mathsf{M}(2,15)$, where $\CG$ does not agree with $C_2$ because $\rho(T)$ has degenerate eigenvalues for fields that are not complex conjugates.} 
\end{table}

The Galois symmetry appears in the induced modular representations of Hecke images, as well as in the definition of Hecke operators for $\Gamma(N)$.
Since the second line of eq\eqref{rep_gen} is merely an alternative form of $\Delta_N^{(p)}$, one can rewrite the Hecke operation as
\begin{align}\begin{split}\label{N_Hecke:define2}
(\mathsf{T}_p f)_i(\tau) &= 
\sum_{j} \rho(\sigma_{p})_{ij}
\sum_{\delta \in \sigma_{\bar p} \circ \Delta_N^{(p)} } f_j(\delta \tau) \\
&= \sum_{j} \rho(\sigma_{p})_{ij}\left(
f_j(p\tau) + \sum_{k} \rho(\sigma_{\bar p})_{jk} \sum_{b=0}^{p-1} f_k \Big(\frac{\tau+bN}{p}\Big)
\right).
\end{split}\end{align}
From the physical point of view, the Hecke operation by $\mathsf{T}_p$ changes the Fourier coefficients, followed by a signed permutation by $\rho(\sigma_p)=G_p^{-1}$. Along the way, the conformal weights in RCFT are multiplied by $p$ modulo $\IZ$.
As shown in \cite{Hecke:2018}, $\mathsf{T}_p f$ transforms in the modular representation
\begin{align}
\rho^{(p)}(\gamma)=f_{N,\bar p}\big( \rho(\gamma) \big)~, \qquad \gamma \in SL(2,\IZ).
\end{align}
The Frobenius map $f_{N,\bar p}$ is a composition of $f_{N, p}$ and $f_{N,\bar p^2}$, both of which have interpretations. 
The action of $f_{N,\bar p^2}$ amounts to conjugation under $G_{p}$, i.e.
\begin{align}\label{f:pbar}
f_{N,\bar p}\big( \rho(\gamma) \big) = 
f_{N,\bar p^2} \circ f_{N,p}\big( \rho(\gamma) \big) =
G_p^{-1}\,f_{N,p}\big( \rho(\gamma) \big) \, G_p .
\end{align}
While the remaining $f_{N, p}$ causes the observed relations between the conformal weights:
\begin{align}\label{field_map}
\tilde\theta_i ~~\mapsto ~~
\theta^{(p)}_i = f_{N,p} (\tilde\theta_i),
\end{align}
or equivalently $h^{(p)}_i \equiv p\, \tilde h_i \pmod{1}$, where $\tilde\theta_i$ and $\tilde h_i$ are respectively the twists and the conformal weights in the effective description to be introduced in Section \ref{sec:eff_picture}. 
The twists $\theta^{(p)}_i $ and $\tilde\theta_i$ are roots of unity of same order, and their associated primary fields share similar statistical (braiding) properties. We will see this essential fact in the full structure of RCFT under Hecke operations.

\subsection{Simple-current reduction of affine algebra}
\label{subsec:simple_current_reduction}
We turn to a number of less familiar RCFT characters and explore the structure  of their corresponding MTCs upon Hecke operations. In condensed matter physics, the (2+1)-dimensional (2+1D) bosonic topological orders are classified by unitary MTCs \cite{Kitaev:anyon,Bonderson:thesis,Rowell:2009}, and simple-current reduction is an important tool in this construction \cite{Wen:2015,Schoutens:2016}. (Various generalizations of unitary MTCs to non-unitary categories also describe 2+1D topological quantum field theories \cite{TQFT_Non-unitary_MTC}. But non-unitary ones do not really have a correspondence with respect to gapped phases of matter.)
A simple current $J$ by definition has a single primary field $J \phi$ appearing in the fusion of $J$ with any primary field $\phi$, thus $J$ permutes the fields by $J \times \phi = J \phi$,  and divides the field content into orbits under the action of $J$. 
Because there are a finite number of primary fields, there exists a smallest positive integer $n$ such that $J^n = I $ in the sense of fusion. This $n$ is called the order of the simple current $J$.
See \cite{Schweigert:thesis} for a general discussion of simple currents.

In $(A_1,k)$ with odd $k$, the spin-$\frac{k}{2}$ primary field $\varphi_{\frac{k}{2}}$ is a simple current with the fusion rules
\begin{align}
\varphi_j \times \varphi_{\frac{k}{2}} = \varphi_{\frac{k}{2}-j}~ ,
\qquad j=0,\frac{1}{2},1, \cdots,\frac{k}{2}.
\end{align}
It maps the half-integer representations onto the integer representations, and vice versa.
The MTC $(A_1,k)_{\frac{1}{2}}$ for $k$ odd consists of the primary fields of integer spin in $(A_1,k)$, and is called the even half of $(A_1,k)$ \cite{Schoutens:2016,Rowell:2009}. We use the notation $\overline{(A_1,k)}$ for the MTC whose twists and modular representation are the complex conjugate of those of $(A_1,k)$. 
With $c=c(A_1,k)\mp 1$,
the MTC is understood as the tensor product
\begin{align}
(A_1,k)= 
\begin{cases} 
(A_1,k)_{\frac{1}{2}} \otimes (A_1,1) ,
\quad &\text{if } k-1\equiv 0  \pmod{4}; \\
(A_1,k)_{\frac{1}{2}} \otimes \overline{(A_1,1)} ,
\quad &\text{if } k+1\equiv 0  \pmod{4}.
\end{cases} 
\end{align}
 
The first nontrivial example is $(A_1,3)_{\frac{1}{2}}$, the integer subset of $(A_1,3)$. It contains the primary fields $\varphi_0$ and $\varphi_1$ with the fusion rules
\begin{align}\label{Fib_fusion_rule}
\varphi_0 \times \varphi_0 = \varphi_0, \qquad
\varphi_0 \times \varphi_1 = \varphi_1, \qquad
\varphi_1 \times \varphi_1= \varphi_0 + \varphi_1,
\end{align}
which are isomorphic to those of the Fibonacci theory.\footnote{The Fibonacci MTC is basically a rank-2 MTC with the fusion rules identical to eq\eqref{Fib_fusion_rule} \cite{Fib_MTC}.}
Moreover, the modular data confirm that $(A_1,3)_{\frac{1}{2}}$ sits in the Fibonacci MTC like $(G_2,1)$. 

Next we study two specific examples, $(A_1,5)_{\frac{1}{2}}$ and $(A_1,7)_{\frac{1}{2}}$. Their modular representations are closely related to the minimal models $\mathsf{M}(2,7)$ and $\mathsf{M}(2,9)$ respectively. We study the Hecke images of their characters and modular representations, as well as the realizations of these Hecke images in VOAs and MTCs.

\subsubsection{Rank three}

The $(A_1,5)_{\frac{1}{2}}$ MTC is realized at central charge
\begin{align}
c \left[(A_1,5)_{\frac{1}{2}} \right] = c \big[(A_1,5) \big] - c \big[(A_1,1) \big]
=15/7 - 1 = 8/7,
\end{align}
and its conformal weights are computed from $(A_1,5)$ as
\begin{align}
h_{j=0}=0,\qquad
h_{j=1}=\frac{2}{7},\qquad
h_{j=2}=\frac{6}{7}.
\end{align}
With this basis ordering, the modular representation $\rho^{(A_1,5)_{\frac{1}{2}}}$ is determined by
\beqs \label{modular_rep:A_5_half}
\begin{align}
\rho^{(A_1,5)_{\frac{1}{2}}}(T) &= \me\Big( -\frac{1}{21} \Big)
\left(
\begin{array}{ccc}
 1 & 0 & 0 \\
 0 & \me\big(\frac{2}{7}\big) & 0 \\ 
 0 & 0 & \me\big(\frac{6}{7}\big) \\
\end{array}
\right),
\\
\rho^{(A_1,5)_{\frac{1}{2}}}(S) &=
\frac{2 \sin \left(\frac{\pi }{7}\right)}{\sqrt{7}}
\left(
\begin{array}{ccc}
 1 & d^2-1 & d \\
 d^2-1 & -d & 1 \\
 d & 1 & 1-d^2 \\
\end{array}
\right),
\end{align}
\eeqs
where $d=2 \cos \left(\frac{\pi }{7}\right)$.  From $\rho^{(A_1,5)_{\frac{1}{2}}}(T)$ we see that the conductor  is $N=21$.

The minimal model $\mathsf{M}(2,7)$ has central charge $c\big[\mathsf{M}(2,7)\big]=-68/7$ and conformal weights
\begin{align}\label{M27:weight}
\Big( h_{1,1} ,h_{3,1} ,h_{5,1}\Big)
= \left(
0, -\frac{3}{7} , -\frac{2}{7} 
\right).
\end{align}
The labeling of conformal weights will be discussed in the next section.
The  conductor is $N'=42$ and the modular representation is given by
\beqs
\begin{align}
\rho^{\mathsf{M}(2,7)}(T) &= 
\me\Big( \frac{17}{42} \Big)
\left(
\begin{array}{ccc}
 1 & 0 & 0 \\
 0 & \me\big(-\frac{3}{7}\big) & 0 \\ 
 0 & 0 & \me\big(-\frac{2}{7}\big) \\
\end{array}
\right), \\
\rho^{\mathsf{M}(2,7)}(S) &= 
\frac{2 \sin\left(\frac{\pi}{7}\right) }{\sqrt{7}} 
\left(
\begin{array}{ccc}
d & 1 & 1-d^2 \\
1 & d^2-1 & d \\
1-d^2 & d & -1 \\
\end{array}
\right).
\end{align}
\eeqs
Hecke images of the $\mathsf{M}(2,7)$ characters were computed and in some cases are vector-valued modular forms that
have appeared in other context,  but they fail to be the characters of a unitary RCFT, because of the existence of negative Fourier coefficients and fusion coefficients \cite{Hecke:2018}. 

Three-dimensional modular representations whose kernels contain congruence subgroups have been classified by Theorem 2 in \cite{eholzerII}. According to the classification, $\rho^{\mathsf{M}(2,7)}$ and $f_{42,-5}\big( \rho^{(A_1,5)_{\frac{1}{2}}} \big)$ differ only by a one-dimensional representation 
\begin{align}
\rho^{\text{1d}}(T)=\me\Big(\frac{1}{6}\Big), 
\qquad
\rho^{\text{1d}}(S)=-1.
\end{align}
This explains why $(A_1,5)_{\frac{1}{2}}$ has fusion rules that are isomorphic to those of $\mathsf{M}(2,7)$. 

The complete set of $\rho^{\mathsf{M}(2,7)}(\sigma_p)$ are provided in \cite{Hecke:2018}.
Explicit computation leads to the following $\rho^{(A_1,5)_{\frac{1}{2}}}(\sigma_p)$ for all $p$, with $1\leq p<21$ and  $\text{gcd}(p,21)=1$. 
\\
When $p=1, 8, 13, 20$,
\begin{align}
\rho^{(A_1,5)_{\frac{1}{2}}}(\sigma_p)=
\mathbb{I}_3~,
\qquad p^2=1 \pmod{21}.
\end{align}
When $p=2, 5, 16, 19$,
\begin{align}
\rho^{(A_1,5)_{\frac{1}{2}}}(\sigma_p)=
\left(
\begin{array}{ccc}
 0 & -1 & 0 \\
 0 & 0 & 1 \\
 -1 & 0 & 0 \\
\end{array}
\right)~,
\qquad p^2=4 \pmod{21}.
\end{align}
When $p=4, 10, 11, 17$,
\begin{align}
\rho^{(A_1,5)_{\frac{1}{2}}}(\sigma_p)=
\left(
\begin{array}{ccc}
 0 & 0 & -1 \\
 -1 & 0 & 0 \\
 0 & 1 & 0 \\
\end{array}
\right)~,
\qquad p^2=16 \pmod{21}.
\end{align}

Among all the $p \in (\mathbb{Z}/N\mathbb{Z})^{\times}$, the series $p=1, 8, 13, 20$ give rise to two inequivalent unitary MTCs that are complex conjugates. 
These $\rho^{(p)}$ have VOA realizations \cite{Wang:2017}, whose characters are not Hecke images of any primitive characters under $\mathsf{T}_p$ though. For $p=8, 13, 20,$ the central charge inferred from the characters is $8p/7$; while for $p=1$ there is no Kac-Moody sub-VOA for $(A_1,5)_{\frac{1}{2}}$ at central charge $8/7$ \cite{Junla:thesis}. Nevertheless there exists a three-character corresponding to $c=22*8/7 \equiv 8/7 \pmod{24}$ and thereby the modular representation $\rho^{(A_1,5)_{\frac{1}{2}}}$.
The case with $c=8*8/7$ is realized as a simple-current reduction of $(A_1,5)\otimes (E_7,1)$. 
However for $p=13,20,22$, the characters associated to these unitary MTCs still lack RCFT interpretations. They are not linked by Hecke operations neither.
When $p^2=4 \text{ or }16 \pmod{21}$, the induced MTCs by $\mathsf{T}_p$ are non-unitary, and there are no Hecke image interpretations.

\subsubsection{Rank four}

We present a similar relation between $\mathsf{M}(2,9)$ and $(A_1,7)_{\frac{1}{2}}$, which have the common conductor $N=36$.
The $(A_1,7)_{\frac{1}{2}}$ MTC has central charge $c \left[ (A_1,7)_{\frac{1}{2}} \right]=10/3$ and twists
\begin{align}\label{A1_7_half:weight}
\big\{ \theta_j \big\}=\big\{ \me(h_{j}) \big\}
=\left\{ 1, 
\me\Big(\frac{2}{9}\Big),
\me\Big(\frac{2}{3}\Big),
\me\Big(\frac{1}{3}\Big) \right\}.
\end{align}
With this ordering of the twists, the modular representation is determined by
\beqs
\begin{align}
\rho^{(A_1,7)_{\frac{1}{2}}}(T) &=
\me\Big(-\frac{5}{36}\Big)
\left(
\begin{array}{cccc}
 1 & 0 & 0 & 0\\
 0 & \me\big(\frac{2}{9}\big) & 0 & 0\\ 
 0 & 0 & \me\big(\frac{2}{3}\big) & 0\\
 0 & 0 & 0 & \me\big(\frac{1}{3}\big)
\end{array}
\right),\\
\rho^{(A_1,7)_{\frac{1}{2}}}(S) &=
\frac{2 \sin \left(\frac{\pi }{9}\right)}{3}
\left(
\begin{array}{cccc}
 1 & r^2-1 & r+1 & r \\
 r^2-1 & 0 & 1-r^2 & r^2-1 \\
 r+1 & 1-r^2 & r & -1 \\
 r & r^2-1 & -1 & -r-1 
\end{array}
\right) ,
\end{align}
\eeqs
where $r=2 \cos \left(\frac{\pi }{9}\right)$.
The minimal model $\mathsf{M}(2,9)$ has central charge $c\big[\mathsf{M}(2,9)\big]=-46/3$ and conformal weights 
\begin{align}\label{M29:weight}
\Big( h_{1,1} ,h_{3,1} ,h_{5,1} ,h_{7,1} \Big)
= \left(
0, -\frac{5}{9}, -\frac{2}{3} , -\frac{1}{3} 
\right).
\end{align}
The modular representation of $\mathsf{M}(2,9)$ is
\beqs\label{rho:M29}
\begin{align}
\rho^{\mathsf{M}(2,9)}(T) &=
\me\Big(\frac{23}{36}\Big)
\left(
\begin{array}{cccc}
 1 & 0 & 0 & 0\\
 0 & \me\big(-\frac{5}{9}\big) & 0 & 0\\ 
 0 & 0 & \me\big(-\frac{2}{3}\big) & 0\\
 0 & 0 & 0 & \me\big(-\frac{1}{3}\big)
\end{array}
\right),
\\
\rho^{\mathsf{M}(2,9)}(S) &=
\frac{2 \sin \left(\frac{\pi }{9}\right)}{3}
\left(
\begin{array}{cccc}
 -r & 1-r^2 & 1 & r+1 \\
 1-r^2 & 0 & r^2-1 & 1-r^2 \\
 1 & r^2-1 & r+1 & r \\
 r+1 & 1-r^2 & r & -1 \\
\end{array}
\right) .
\end{align}
\eeqs
Note that $\rho^{\mathsf{M}(2,9)} $ differs from $f_{36,-7}\left( \rho^{(A_1,7)_{\frac{1}{2}}} \right)$ by a one-dimensional representation 
\begin{align}
\rho^{\text{1d}}(T)=\me\Big(\frac{-1}{3}\Big), 
\qquad
\rho^{\text{1d}}(S)=1,
\end{align}
yielding the same fusion rules.

Explicit computations lead to the following $\rho(\sigma_p)$ for all $p$, with $1\leq p<36$ and  $\text{gcd}(p,36)=1$. The bases are ordered following eq\eqref{M29:weight} and \eqref{A1_7_half:weight} respectively. \\
When $p=1,35$, ($p^2=1 \text{ mod }36$)
\begin{align}
\rho^{\mathsf{M}(2,9)}(\sigma_p)= 
\rho^{(A_1,7)_{\frac{1}{2}}}(\sigma_p)= \mathbb{I}_4  ~.
\end{align}
When $p=17,19$, ($p^2=1 \text{ mod }36$)
\begin{align}
\rho^{\mathsf{M}(2,9)}(\sigma_p)= 
\rho^{(A_1,7)_{\frac{1}{2}}}(\sigma_p)= 
- \mathbb{I}_4  ~.
\end{align}
When $p=5,31$, ($p^2=25 \text{ mod }36$)
\begin{align}
\rho^{\mathsf{M}(2,9)}(\sigma_p)=
\rho^{(A_1,7)_{\frac{1}{2}}}(\sigma_p)=
\left(
\begin{array}{cccc}
 0 & 0 & 0 & -1 \\
 0 & -1 & 0 & 0 \\
 1 & 0 & 0 & 0 \\
 0 & 0 & 1 & 0 \\
\end{array}
\right).
\end{align}
When $p=13,23$, ($p^2=25 \text{ mod }36$)
\begin{align}
\rho^{\mathsf{M}(2,9)}(\sigma_p)=
\rho^{(A_1,7)_{\frac{1}{2}}}(\sigma_p)= -
\left(
\begin{array}{cccc}
 0 & 0 & 0 & -1 \\
 0 & -1 & 0 & 0 \\
 1 & 0 & 0 & 0 \\
 0 & 0 & 1 & 0 \\
\end{array}
\right).
\end{align}
When $p=7,29$, ($p^2=49 \text{ mod }36$)
\begin{align}
\rho^{\mathsf{M}(2,9)}(\sigma_{p})=
\rho^{(A_1,7)_{\frac{1}{2}}}(\sigma_p)= 
\left(
\begin{array}{cccc}
 0 & 0 & 1 & 0 \\
 0 & -1 & 0 & 0 \\
 0 & 0 & 0 & 1 \\
 -1 & 0 & 0 & 0 \\
\end{array}
\right).
\end{align}
When $p=11,25$, ($p^2=49 \text{ mod }36$)
\begin{align}
\rho^{\mathsf{M}(2,9)}(\sigma_{p})= 
\rho^{(A_1,7)_{\frac{1}{2}}}(\sigma_p)= -
\left(
\begin{array}{cccc}
 0 & 0 & 1 & 0 \\
 0 & -1 & 0 & 0 \\
 0 & 0 & 0 & 1 \\
 -1 & 0 & 0 & 0 \\
\end{array}
\right).
\end{align}
Again, the observation that $\rho^{\mathsf{M}(2,9)}(\sigma_{p})= 
\rho^{(A_1,7)_{\frac{1}{2}}}(\sigma_p)$ is explained by the underlying Galois symmetry between the two MTCs.

Some Hecke images of $\chi^{\mathsf{M}(2,9)}$ give the characters of affine Lie algebras:
\begin{align}
\mathsf{T}_7 \chi^{\mathsf{M}(2,9)} &= \chi^{(G_2,2)}, \\
\mathsf{T}_{29} \chi^{\mathsf{M}(2,9)} &= \chi^{(C_5,3) \otimes (A_1,1)}.
\end{align}
These four-character theories 
are listed in Table 3 of \cite{Gaberdiel:coset}. Their bilinear form 
\begin{align}
\mathsf{T}_7 \chi^{\mathsf{M}(2,9)} \cdot \mathsf{T}_{29} \chi^{\mathsf{M}(2,9)}
=J(\tau)+72
\end{align}
reproduces the partition function of  No. 21 in Schelleken's list of $c=24$ meromorphic CFTs \cite{Schellekens:c24}, where
\begin{align}
J(\tau)= q^{-1}+196884 q + 21493760 q^2 + 864299970 q^3 +\cdots
\end{align}
is the modular $J$ function.

$(A_1,7)_{\frac{1}{2}}$ has a VOA realization as the simple-current reduction of $(A_1,7)\otimes (A_1,1)$.
Its characters are constructed as
\begin{align}
\chi^{(A_1,7)_{\frac{1}{2}}}_j (\tau)=
\chi^{(A_1,7)}_j(\tau) \chi^{(A_1,1)}_0(\tau) +
\chi^{(A_1,7)}_{\frac{7}{2}-j}(\tau) 
\chi^{(A_1,1)}_{ \frac{1}{2} }(\tau),
\end{align}
where the subscript $j=0,1,2,3$ stands for the spin-$j$ representation of $SU(2)$. The characters afford the $q$-series expansions
\beqs
\begin{align}
\chi^{(A_1,7)_{\frac{1}{2}}}_0(\tau) &= q^{-\frac{5}{36}} (1 + 6 q + 38 q^2 + 112 q^3 + 347 q^4 +\cdots) ,\\
\chi^{(A_1,7)_{\frac{1}{2}}}_1(\tau) &= q^{\frac{1}{12}} (3 + 30 q + 114 q^2 + 384 q^3 + 1065 q^4 +\cdots) ,\\
\chi^{(A_1,7)_{\frac{1}{2}}}_2(\tau) &= q^{\frac{19}{36}} (13 + 62 q + 230 q^2 + 692 q^3 + 1874 q^4 +\cdots) ,\\
\chi^{(A_1,7)_{\frac{1}{2}}}_3(\tau) &= q^{\frac{7}{36}} (4 + 23 q + 102 q^2 + 319 q^3 + 886 q^4 +\cdots) .
\end{align}
\eeqs
In principle, the Hecke images of the $(A_1,7)_{\frac{1}{2}}$ characters can be calculated by the standard algorithm.
Various Galois-conjugate representations of $(A_1,7)_{\frac{1}{2}}$ are listed in Table 11 of \cite{eholzerII}, which summarizes four-dimensional simple strongly-modular fusion algebras up to one-dimensional modular representations.

Moreover, the MTCs of $(A_1,7)_{\frac{1}{2}}$ and $(G_2,2)$ are complex conjugate \cite{Schoutens:2016}. Their characters make up the bilinear form
\begin{align}
\chi^{(A_1,7)_{\frac{1}{2}}} (\tau) \cdot \chi^{(G_2,2)} (\tau) = j(\tau)^{1/3} ,
\end{align}
where $ j(\tau) = J(\tau) + 744$ is the $j$-invariant.

\subsection{MTCs of higher rank}
\label{subsec:higher_rank}
The previous examples suggest a connection between $\mathsf{M}(2,k+2)$ and $(A_1,k)_{\frac{1}{2}}$, since their fusion rules are isomorphic. 
This connection offers a series of examples that Galois conjugations convert non-unitary RCFTs to unitary ones, and vice versa.
Moreover, both theories are related to critical behaviors of chains of antiferromagnetically coupled anyons as pointed out in \cite{Ardonne}.

The primary fields in $\mathsf{M}(2,k+2)$ are denoted by $\phi_{(u,1)}$ with $u$ an odd integer satisfying $1\leq u <k+2$. They respect the fusion rules
\begin{align}\label{fusion:M}
\phi_{(u_1,1)} \times \phi_{(u_2,1)} = 
\sum^{u_{\rm max} }_{\substack{u=1+| u_1-u_2 | \\ u \text { odd} } } \phi_{(u,1)} ~,
\end{align}
where $u_{\rm max } = \min(u_1+u_2-1,2k+3-u_1-u_2)$ \cite{Francesco:2012}.
The fusion rules of $(A_1,k)$ resemble the compositions of $SU(2)$ angular momenta, namely
\begin{align}\label{fusion:A}
\varphi_{l_1} \times \varphi_{l_2} = 
\sum^{l_{\rm max}}_{l=1+| l_1-l_2 | } \varphi_l ~,
\end{align}
where $l_{\rm max}=\min(l_1+l_2-1,2k+3-l_1-l_2)$ \cite{Ardonne}. The label $l=2j+1$ is the number of states for integral spin-$j$, and is odd in $(A_1,k)_{\frac{1}{2}}$. Evidently, the fusion rules of $\mathsf{M}(2,k+2)$ and $(A_1,k)_{\frac{1}{2}}$ are isomorphic with the identification $\phi_{(l,1)} \sim \varphi_l$.

More fundamentally, the isomorphism of fusion rules stems from the Galois symmetry between $\mathsf{M}(2,k+2)$ and $(A_1,k)_{\frac{1}{2}}$.
To conduct a general analysis, we set $N=24(k+2)$, which is an integral multiple of both conductors. All the modular data are in the number field $\IQ[\xi_N]$.
We claim that $\rho^{\mathsf{M}(2,k+2)} $ differs from the $(-k)$-th Galois conjugate of $ \rho^{(A_1,k)_{\frac{1}{2}}} $ by the one-dimensional representation $\rho^{\text{1d}}$, 
where
\begin{align}
\rho^{\text{1d}}(T)=\me\Big(\frac{t}{6}\Big), 
\qquad
\rho^{\text{1d}}(S)=(-1)^t,
\end{align}
if $k=4t+1$, or
\begin{align}
\rho^{\text{1d}}(T)=\me\Big(-\frac{t}{6}\Big), 
\qquad
\rho^{\text{1d}}(S)=(-1)^t,
\end{align}
if $k=4t-1$ with $t\in \IZ_+$. 
The proof of this assertion is left to Appendix \ref{appen:Galois}.
For the moment, the physical meaning of $\rho^{\text{1d}}$ is unclear here. These are 1D representations of a modular fusion algebra, but are not representations appearing in any known RCFT. 1D representations of modular fusion algebras are possible for central charge $c$ a multiple of $4$ \cite{eholzerII}, but RCFT/VOA realizations are only known for $c$ a multiple of $8$. 
When $c\equiv 0 \pmod{8}$, a RCFT such as affine $E_8$ at level one corresponds to the trivial MTC.

In summary, the $\mathsf{M}(2,k+2)$ characters are known for general odd $k$ and their Hecke images are computable. Though not unitary, $\mathsf{M}(2,k+2)$ has a unitarization realized as $(A_1,k)_{\frac{1}{2}}$ by Galois symmetry.
When $k\equiv -1 \pmod{4}$, the $(A_1,k)_{\frac{1}{2}}$ characters are constructed in an analogous way to those for $(A_1,7)_{\frac{1}{2}}$, and can be acted on by Hecke operators.
Although there is no standard way to realize $(A_1,k)_{\frac{1}{2}}$ via a VOA when $k\equiv 1 \pmod{4}$, Hecke operators can still be implemented on the level of character. 


As mentioned earlier, some MTCs have ranks greater than four and involve complex-conjugate  pairs of primaries. In such cases, we may identify each pair of complex primaries and reduce the modular representation to a smaller dimensionality, before acting with the Hecke operators. For example, affine $SU(3)$ at level 1 has two primaries that create fields in the $3$ and $\bar 3$ of $SU(3)$, but since
they have the same character one can construct a two-dimensional representation of the modular group given by the modular transformation
of the vacuum character and one of these characters. 
As a more complicated example of this technique, we start with the six-character $\psi$ associated to a special RCFT, which has $c=8/5$ and $N=15$. Let $M_I$ be the MTC of this RCFT.
\beqs
\begin{align}
\psi_0 (\tau) &=q^{-\frac{1}{15}}(1+ 4q+ 8q^2+ 20q^3+ 37q^4 +\cdots), \\
\psi_{\frac{2}{15}} (\tau) &=q^{\frac{1}{15}}(1+ 2q+ 7q^2+ 12q^3+ 26q^4 +\cdots), \\
\psi_{\frac{2}{15}}^* (\tau) &= \psi_{\frac{2}{15}} (\tau), \\
\psi_{\frac{4}{5}} (\tau) &=q^{\frac{11}{15}}(3+ 4q+ 10q^2+ 20q^3+ 38q^4 +\cdots), \\
\psi_{\frac{1}{3}} (\tau) &=q^{\frac{4}{15}}(2+ 5q+ 12q^2+ 23q^3+ 46q^4 +\cdots), \\
\psi_{\frac{1}{3}}^* (\tau) &= \psi_{\frac{1}{3}} (\tau) ,
\end{align}
\eeqs
where the sub-index of $\psi$ refers to the conformal weight. 
This RCFT has two pairs of complex primaries, which are of conformal weights $2/15$ and $1/3$ respectively. 
It is constructed as an intermediate vertex sub-algebra like those in \cite{kawasetsu,sextonion}. 
Moreover, its modular $T$ matrix is of odd order, though the orders of $\rho(T)$ tend to be even in generic RCFTs.
The components of 
\begin{align}
\psi=\Big( \psi_0,\psi_{\frac{2}{15}},\psi_{\frac{4}{5}},\psi_{\frac{1}{3}} \Big)
\end{align}
are solutions to a fourth-order modular linear differential equation (MLDE), and are closed under the $SL(2,\IZ)$ transformations since the MLDE is modular invariant. The differential equation involves three free parameters $\mu_1,\mu_2,\mu_3$, and takes the form
\begin{align}\label{MLDE:order4}
\big( \CD^4+\mu_1 E_4 \CD^2+ \mu_2 E_6 \CD +\mu_3 E_8 \big) f = 0,
\end{align}
where $\CD=\md/\md\tau-\frac{1}{6}\mi\pi k E_2$ is the Serre derivative acting on weight-$k$ modular forms, and $E_j$ is the Eisenstein series of weight $j$ \cite{arikeII}.
Given an $n$th-order MLDE, the Wronskians are constructed out of the $n$ linearly independent solutions as
\be \label{Wronskian:def}
W_k =  \begin{vmatrix} f_1 & f_2 & \cdots & f_n \\
                                                   {\cal D} f_1 & {\cal D} f_2 & \cdots & {\cal D} f_n \\
                                                   \vdots & \vdots &      & \vdots \\
                                                   {\cal D}^{k-1} f_1 & {\cal D}^{k-1} f_2 & \cdots & {\cal D}^{k-1} f_n \\
                                                    {\cal D}^{k+1} f_1 & {\cal D}^{k+1} f_2 & \cdots  & {\cal D}^{k+1} f_n \\
                                                    \vdots & \vdots &      & \vdots \\
                                                     {\cal D}^{n} f_1 & {\cal D}^{n} f_2 & \cdots  & {\cal D}^{n} f_n
                                                     \end{vmatrix} \, .
 \ee
We denote by $\ell(W)/6$ the number of zeros in the Wronskian $W=W_n$ \cite{Mathur:1988gt}. The form of eq\eqref{MLDE:order4} implies $\ell(W)=0$ here.
The $(A_2,1)$ MTC is the tensor product of $M_I$ and the Yang-Lee model, followed by a simple-current reduction. 
Neither $M_I$ nor $\psi$ is listed in the VOA encyclopedia \cite{VOA_database} because $M_I$ is non-unitary. ($c=8/5$ should be understood as the effective central charge to be introduced later.) We anticipate a connection to the three-state Potts model due to resemblance of the field contents as well as the modular representations. Inspired by Consequence 4 in \cite{Gannon:nonunitary}, we predict that the VOA of $\psi$ contains the $\CW_3$ algebra.

As a vector-valued modular form, $\psi$ has Hecke images that are characters of affine Lie algebras:
\begin{align}
\mathsf{T}_2 \psi &= \chi^{(A_2,2)} ,\\
\mathsf{T}_{13} \psi &= \chi^{(F_4,6)} .
\end{align}
They are also obtained by solving eq\eqref{MLDE:order4} \cite{Gaberdiel:coset}.
When treated as six-character theories, $(A_2,2)$ and $(F_4,6)$ each have two complex-conjugate fields, in accord with the preimage $\psi$.
Their MTCs are complex conjugates, and the bilinear form of their characters is modular invariant.
\begin{align}\begin{split}\label{bilinear:2_13}
&\chi^{(A_2,2)}_0 \chi^{(F_4,6)}_0
+2\chi^{(A_2,2)}_{\frac{4}{15}} \chi^{(F_4,6)}_{\frac{26}{15}}
+\chi^{(A_2,2)}_{\frac{3}{5}} \chi^{(F_4,6)}_{\frac{7}{5}}
+2\chi^{(A_2,2)}_{\frac{2}{3}} \chi^{(F_4,6)}_{\frac{4}{3}} \\
=& q^{-1} + 60 + 196884 q + 21493760 q^2 +\cdots
\equiv J(\tau)+60.
\end{split}\end{align}
The multiplicity 2 accounts for the complex primaries and is crucial to attain the modular invariance.
This bilinear form produces Schellekens No. 14 \cite{Schellekens:c24}.
The Hecke image of $\psi$ under $\mathsf{T}_{14}$ yields  positive $q$-series, which can also be constructed by acting with $\mathsf{T}_{7}$ on $\chi^{(A_2,2)}=\mathsf{T}_2 \psi$ since $\mathsf{T}_{mn}=\mathsf{T}_{m}\mathsf{T}_{n}$ for $\text{gcd}(m,n)=1$.
\beqs
\begin{align}
\big(\mathsf{T}_{14} \psi\big)_0 &= q^{-\frac{14}{15}}(1+ 56q+ 87836q^2+ 7358176q^3 +\cdots), \\ 
\big(\mathsf{T}_{14} \psi\big)_{\frac{28}{15}} &= q^{\frac{14}{15}}(26730+ 2694384q+ 99032220q^2 +\cdots),\\
\big(\mathsf{T}_{14} \psi\big)^*_{\frac{28}{15}} &= \big(\mathsf{T}_{14} \psi\big)_{\frac{28}{15}},\\
\big(\mathsf{T}_{14} \psi\big)_{\frac{6}{5}} &= q^{\frac{4}{15}}(308+ 147280q+ 9692893q^2 +\cdots),\\
\big(\mathsf{T}_{14} \psi\big)_{\frac{5}{3}} &= q^{\frac{11}{15}}(13608+ 1927233q+ 82069848q^2 +\cdots),\\
\big(\mathsf{T}_{14} \psi\big)^*_{\frac{5}{3}} &= \big(\mathsf{T}_{14} \psi\big)_{\frac{5}{3}}.
\end{align}
\eeqs
$\psi$ and $\mathsf{T}_{14} \psi$ correspond to complex-conjugate MTCs, which are non-unitary. Their bilinear form gives the identical modular invariant $J(\tau)+60$ as eq\eqref{bilinear:2_13}.


\section{Picture of Effective Central Charge}
\label{sec:eff_picture}
We start with the characters $\chi$ of a RCFT (unitary or non-unitary) with effective central charge $c_{\rm eff}$. If $\mathsf{T}_p \chi$ are also the characters of a RCFT, then it follows from the formula for the Hecke transform that the central charge is
\begin{align}
c^{(p)}=p\,c_{\rm eff},
\end{align}
as long as the criteria of unitarity are met \cite{Hecke:2018}. 
Moreover, the conformal weights also change upon Hecke operations, as seen in eq\eqref{field_map}.
However the Hecke operation does not necessarily map the vacuum character of the original to the vacuum character of the new
theory.  To clarify the Hecke operation, we seek a systematic approach to the picture of effective central charge.
This method aligns the fields in the Hecke image with the initial theory by similar statistics, and helps to locate the vacuum entry.

We use the notation that if $X$ refers to a quantity in the initial theory, then $\tilde X$ and $X^{(p)}$ stand for the counterparts in the effective description and the Hecke image under $\mathsf{T}_p$ respectively.

\subsection{Unified method}
\label{subsec:unified_method}
The Virasoro minimal models were briefly mentioned in Section \ref{subsec:Galois_Permu}. They have well-known characters, and these characters provide an interesting class of vector-valued modular forms that can be acted on by Hecke operators. The minimal model $\mathsf{M}(p_1, p_2)$ has central charge
\begin{align}\label{c:p1p2}
c \big[\mathsf{M}(p_1,p_2)\big] = 1- 6\frac{(p_1-p_2)^2}{p_1p_2}
\end{align}
and conformal weights
\begin{align}\label{h:p1p2}
h_{r,s}=\frac{(p_1 r-p_2 s)^2-(p_1-p_2)^2}{4p_1 p_2}
\end{align}
for the primary fields labeled by $(r,s)$ with $0<r<p_2$, $0<s<p_1$. 
In a non-unitary minimal model, the central charge and the conformal weights can be negative. To provide a general analysis Gannon defines the minimal primary $\mathit{o}=(r_{\mathit o},s_{\mathit o})$ to be the primary field of lowest conformal weight, which corresponds to the unique $(r,s)\in \CI$ obeying $p_1 r- p_2 s=1$ \cite{Gannon:nonunitary}.
He also shows that $\mathit{o}$ has a positive $\rho(S)$ column. 
The effective central charge $c_{\text{eff}}$ and the shifted conformal weights $\mathfrak{h}$ are defined so that the character of $\mathit{o}$ has leading singularity $q^{-c_{\text{eff}}/24}$ while other primaries have $q^{\mathfrak{h}-c_{\rm eff}/24}$ as $q\rightarrow 0$. In what follows we denote by $\tilde{\mathsf{M}} (p_1, p_2)$ the effective description of $\mathsf{M}(p_1, p_2)$.
For the minimal model $\mathsf{M}(p_1,p_2)$, one has
\begin{align}\label{c:eff}
c_{\text{eff}} \big[\mathsf{M}(p_1,p_2)\big] 
\equiv c \big[\mathsf{\tilde M}(p_1,p_2)\big]  = 1- \frac{6}{p_1 p_2}
\end{align}
and the shifted conformal weights
\begin{align}\label{h:eff_shift}
\mathfrak{h}_{r,s}=\frac{(p_1 r-p_2 s)^2-1}{4p_1 p_2} ,\qquad (r,s)\in \CI.
\end{align}
$\mathfrak{h}_{r,s}$ is a mere constant shift from $h_{r,s}$, and should not be confused with the effective conformal weights to be presented later.
The conductor remains the same in this description.

We can extend the analysis for minimal models to generic RCFTs.
There are two generic ways to find symmetric $S$ matrices which diagonalize the fusion coefficients $N_i$: simple currents and the Galois symmetry. Under either of them  the symmetry condition of $\rho(S)$ is preserved. There exists a unique chiral primary $\mathit{o}$ called the minimal primary, which has the lowest conformal weight in the RCFT \cite{Gannon:nonunitary}.
Let $c_{\text{eff}}$ still denote the effective central charge. As always, the character of the $\mathit{o}$ primary has the leading term $q^{-c_{\text{eff}}/24}$.
By Gannon's definition \cite{Gannon:nonunitary}, a RCFT is said to have the Galois shuffle (GS) property if there is a simple current $J_{\mathit o}$ (possibly the identity) and a Galois automorphism $\sigma_{\mathit o}$ (possibly the identity), such that the precise relationship 
\begin{align}
\mathit{o}=J_{\mathit o} \times \sigma_{\mathit o} 0 
\end{align}
holds, where $0$ is the vacuum primary. The right hand side is understood as the fusion of the fields $J_{\mathit o}$ and $\sigma_{\mathit o} 0$.
Moreover, $J_{\mathit o}$ is of order 1 or 2 (so $4h_{J_{\mathit o}} \in \IZ$). The GS property obviously holds for unitary theories.
Gannon proves that the GS property is possessed by all $\CW_N$ minimal models, in particular $\CW_2$ also known as the Virasoro minimal models.\footnote{Akin to the Virasoro minimal models, $\CW_N$ minimal models are generated by fields of conformal weight $N$ \cite{Zamolodchikov:cft1985,Fateev:Zn}.}
There are modular representations that do not obey the GS property, for instance the one-dimensional representation $\rho(T)=-1$, $\rho(S)=-1$.

A simple current $J$ permutes the fields by the fusion rule $J \times a = J a$, where there is only one term on the right hand side.
In Section \ref{subsec:simple_current_reduction} we have seen another usage of the simple current, where it reduces the structure of affine algebras. In this section we investigate its role in the effective description of RCFT.
Invoke the property of simple currents 
\begin{align}\label{Q_J:def}
\rho(S)_{Ja,b}= \me\big[Q_J(b)\big] \rho(S)_{a,b}~,
\end{align}
where $Q_J(b)$ is the monodromy charge of the field $b$ under the current $J$ \cite{Schweigert:thesis}. If $J$ is of order $n$, the monodromy charge $Q_J(b) \in \frac{1}{n} \IZ$. In this paper, $Q_J(b)$ is a half-integer and thus $\me\big[Q_J(b)\big] = \pm 1$. The positive column of the minimal primary requires $Q_J(\mathit{o})\in\IZ$ for all simple currents $J$, which boils down to $Q_J(0)\in\IZ$ in unitary theories. 
The monodromy charges are determined by the conformal weights of the fields on the simple current orbit
\begin{align}\label{h_Q}
h_a+h_J-h_{Ja}\equiv Q_J(a)-Q_J(0)  \pmod{1}
\end{align}
\cite{Gannon:nonunitary}. Unlike in the unitary theories, $Q_J(0)$ can be a half-integer when the theory is non-unitary. This yields modifications of the selection rule applying to the unitary theories
\begin{align}\label{selection_rule}
Q_J(a)+Q_J(b)\equiv Q_J(c)+Q_J(0) \pmod{1}
\end{align}
if ${}_0{N_{ab}}^c \neq 0$. 
Apart from the above constraint on fusion rules, the property eq\eqref{Q_J:def} of simple currents demands ${}_0{N_{Ji,J^{-1}j}}^k = {}_0{N_{ij}}^k$. 
The existing fusion rule eq\eqref{fusion_rule} then implies another:
\be\label{fusion_rule:JJ}
\phi_{Ji} \times \phi_{J^{-1}j} = \sum_k {}_0{N_{ij}}^k \, \phi_k .
\ee

The other element of the GS property is the Galois automorphism $\sigma_{\mathit o}$, which is chosen to be the permutation $\pi_{\ell}$ of fields labeled by some $\ell \in (\mathbb{Z}/N\mathbb{Z})^{\times}$.
Recall that $ (G_{\ell})_{i,j} =\varepsilon_{\ell}(i)\,\delta_{\pi_{\ell} (i),j}$.
For instance in the Virasoro minimal models, it permutes the primary fields according to
\begin{align}
\pi_{\ell}: ~ (r,s) \rightarrow (\ell r, \ell s).
\end{align}
This is obvious from shuffling the modular $T$ matrix \eqref{T_l^2 entry} and is also inferred implicitly \cite{Gannon:nonunitary}. 
It is convenient to require $\ell \in (\mathbb{Z}/8p_1p_2\mathbb{Z})^{\times}$ as well, hence $\ell$ is odd.
On the level of modular representations, the permutation $\pi_{\ell}$ gives rise to the inner automorphism $f_{N,\ell^2}$ on $\IQ[\xi_N]$, namely 
\begin{align}
f_{N,\ell^2}\big( \rho(\gamma) \big)= G_{\ell} \, \rho(\gamma) \, G_{\ell}^{-1}~.
\end{align}
The GS property implies the relation
\begin{align}\label{GS_property}
-\frac{c_{\rm eff}}{24} \equiv  -\frac{c}{24}\ell^2 +h_{J_{\mathit o}} 
\equiv \left( -\frac{c}{24} +h_{J_{\mathit o}} \right)\ell^2 \pmod{1}
\end{align}
in the $\rho(T)$ entries, where the second congruence comes from the fact that $4h_{J_{\mathit o}} \in \IZ$. We call this relation the GS equation.
Both $J_{\mathit o}$ and $\sigma_{\mathit o}$ yield signed permutations of the characters. There are only a finite number of quadratic residues 
\begin{align}
\ell^2 \in \big[(\mathbb{Z}/N\mathbb{Z})^{\times}\big]^2
\end{align}
that validate the GS property. The choice of $\ell$ is not unique for each quadratic residue. Once the inner automorphism by suitable $\ell^2$ has been chosen, the simple current $J_{\mathit o}$ is uniquely determined. 
In what follows we omit the subscript of $J_{\mathit o}$ and write the Galois automorphism $\sigma_{\mathit o}$ as $\pi_{\ell}$ with explicit dependence on $\ell$.

In the effective description, the modular representation reads
\begin{align}
\tilde \rho(\gamma)
=f_{N,\ell^2}\big( P \, \rho(\gamma)\, P^{-1} \big) 
= P \, f_{N,\ell^2}\big( \rho(\gamma) \big) \, P^{-1}
=PG_{\ell} \, \rho(\gamma) \,(PG_{\ell})^{-1},
\end{align}
where $P$ is the permutation matrix $P_{ab}=\delta_{Ja,b}$. 
The orthogonal matrices $P$ and $G_{\ell}$ commute, implying that they are different types of permutation.
Hence, one can write $\tilde \rho(\gamma)$ in the form of inner automorphism
\begin{align}
f_{N,\ell^2}\big( P \, \rho(\gamma)\, P^{-1} \big) 
=G_{\ell} \big( P \, \rho(\gamma)\, P^{-1} \big) G_{\ell}^{-1} ,
\end{align}
and regard $P \, \rho(\gamma)\, P^{-1}$ as a new modular matrix.
In terms of the $SL(2,\IZ)$ generators, the matrix elements are
\beqs\label{rho:tilde}
\begin{align}
\tilde \rho(T)_{ab}&= f_{N,\ell^2} \big( \rho(T)_{Ja,Jb} \big)
=\delta_{ab}\, \big(\rho(T)_{Ja,Ja}\big)^{\ell^2}~, \\
\tilde \rho(S)_{ab}&=  f_{N,\ell^2} \big( \rho(S)_{Ja,Jb} \big)
=\me\big[\ell^2 Q_J(Jb)\big]\,
\epsilon_{\ell^2}(Jb) \, \rho(S)_{a,\pi_{\ell^2}Jb}~.
\end{align}
\eeqs
As such, $\tilde \rho(T)$ is obtained by shuffling the diagonal elements of $\rho(T)$.
The effective conformal weights $\tilde h_a$ are inferred from $\tilde \rho(T)$:
\begin{align}\begin{split}\label{h:tilde}
\tilde h_a -\frac{c_{\rm eff}}{24} &\equiv \ell^2\Big(h_{Ja}-\frac{c}{24}\Big) \pmod{1} \\
\Rightarrow \qquad
\tilde h_a &\equiv \ell^2 ( h_{Ja}-h_J ) \pmod{1}.
\end{split}\end{align}
$\tilde h_a$ is not the shifted conformal weight $\mathfrak{h}_a$ in eq\eqref{h:eff_shift} for any $a\in\CI$. Instead, one deduces from the shuffling rule $\tilde \rho(T)_{aa} = \rho(T)_{\pi_{\ell} J a, \pi_{\ell} J a}$ that
\begin{align}\begin{split}\label{h:eff_shift2}
\me\left( \tilde h_a -\frac{c_{\rm eff}}{24} \right)
&=\me\left(h_{\pi_{\ell}Ja}-\frac{c}{24}\right)
=\me\left( \mathfrak{h}_{\pi_{\ell}Ja} -\frac{c_{\rm eff}}{24} \right) \\
\Rightarrow \qquad \tilde h_a &\equiv \mathfrak{h}_{\pi_{\ell}Ja} \pmod{1}.
\end{split}\end{align}
We anticipate that $\tilde \theta_a$ and $\theta_a$ are roots of unity of the same order.
The first row/column of $\tilde\rho(S)$ need not be positive, because $\tilde\rho(S)$ does not necessarily transform positive characters as in RCFT. It will be seen shortly that the form of $\tilde\rho(S)$ enables non-negative fusion rules. 

Based on the GS property, Gannon proposed a method called ``unitarization'', which converts RCFTs to unitary ones with identical fusion rules \cite{Gannon:nonunitary}. 
Given a RCFT with central charge $c$, its unitarization usually has central charge that is an integral multiple of $c$. In this case, the unitarization is essentially equivalent to the method of Hecke operation, where the unitarization focuses on the MTC aspect while the Hecke operation deals with the characters. Without the integral relation of central charges, there is no interpretation for the unitarization in terms of the Hecke image.

Our approach differs from Gannon's in that the effective description exploits the GS property but does not unitarize the initial theory.
There could be several effective descriptions $(\ell^2,J)$ with distinct representations $\tilde \rho(\gamma)$, but they correspond to the unique $c_{\rm eff}$. By the form of $\tilde \rho(\gamma)$, the effective description does not alter the conductor.
Moreover,  $c_{\rm eff}$ are known for the Virasoro minimal models, which allows us to solve $(\ell^2,J)$ in the GS equations. 

A crucial subset of Hecke operators are $\mathsf{T}_p$ with ${\bar p}^2 \equiv \ell^2 \pmod{N}$, where $(\ell^2,J)$ is an effective description. In this case, the Hecke operation $\mathsf{T}_p$ comprises two steps as in eq\eqref{N_Hecke:define2}.
The signed permutation $\rho(\sigma_p)=G_p^{-1}$ implements the same transformation as the inner automorphism $f_{N,\ell^2}$ does in the effective description. 
Like in the effective description, the conductor stays invariant under the Hecke operation.
As we will see, the fusion rules do not change under the joint action by $J$ and $\pi_{\ell}$, with the same ordering in the representation matrix. In the rest of this section, we impose the constraint ${\bar p}^2 \equiv \ell^2 \pmod{N}$ so as to get unitary theories under $\mathsf{T}_p$.

\subsection{Fusion rules of Hecke image}
\label{subsec:FusionRules}
In this subsection we discuss the fusion rules for the Hecke image which are related to  the couplings of the primary fields in the Hecke image. The derived fusion rules are one physical implication of Hecke relations and pave the way for computing the duality properties algebraically.

The Hecke operation $\mathsf{T}_p$ takes the characters of a RCFT with effective central charge $c_{\rm eff}$ to their image characters, which may be characters of a RCFT which has central charge $c^{(p)}=p\,c_{\rm eff}$. The modular representations of the two theories are related by the Frobenius map $f_{N,\bar p}$.
It might seem that the initial theory and its Hecke image have same fusion rules, since the Frobenius map acts trivially on the integral fusion coefficients. 
However this reasoning is not accurate, because the fusion coefficients depend on the vacuum row as shown in the Verlinde formula. 
Though the Hecke image has the modular $S$ matrix $\rho^{(p)}(S) = f_{N,\bar p}\big(\rho(S)\big)$,
$$\left\{ f_{N,\bar p}\big(\rho(S)_{0i}\big) \,\big|\, i \in \CI \right\}$$ 
is no longer the vacuum row as the primaries have been shuffled.

Thus to study fusion rules of the Hecke image we should ensure that the field content is properly aligned so that we can identify the new vacuum character. 
To do so, we first translate the initial modular data to the picture of effective central charge.
With the representation $\tilde\rho$ defined in eq\eqref{rho:tilde}, we compute the fusion rules ${}_0{\tilde N_{ab}}^{~~~c}$ directly.
\begin{align}\begin{split}
\label{FusionRules:same1}
{}_0{\tilde N_{ab}}^{~~~c} &= \sum_m \frac{\tilde\rho(S)_{a,m} \tilde\rho(S)_{b,m} \tilde\rho(S)^{-1}_{c,m}}{ \tilde\rho(S)_{0,m}} 
\\
&= f_{N,\ell^2} \left( \sum_m \frac{\rho(S)_{a,m}\rho(S)_{b,m}\rho(S)^{-1}_{c,m}}{\rho(S)_{0,m}} 
\, \me\big[ Q_J(a)+Q_J(b)-Q_J(c)-Q_J(0)\big] \right) \\
&= f_{N,\ell^2}\big( {}_0{N_{ab}}^c \big) 
= {{}_0 N_{ab}}^c ,
\end{split}\end{align}
where the selection rule eq\eqref{selection_rule} is used. Hence, $\tilde\rho$ leads to the identical fusion rules as in the initial theory, which are of course non-negative. There are not necessarily physical fields that give rise to the representation $\tilde\rho$. 
In fact, the $q$-series that transform under $\tilde\rho$ are the initial characters with a signed permutation, and could have negative Fourier coefficients.

Being brought to the effective picture, it remains to multiply the effective conformal weights by $p$ along with changing the Fourier coefficients. 
It causes an action of $f_{N,p}$ on the modular representation, rendering the fusion rules invariant. The fusion rule in the image theory is
\begin{align}\begin{split}
\label{FusionRules:same2}
{}_0{N_{ab}}^{c \, (p)} 
&= \sum_m \frac{f_{N,p} \big(\tilde\rho(S)\big)_{am} f_{N,p} \big(\tilde\rho(S)\big)_{bm} f_{N,p} \big(\tilde\rho(S)^{-1} \big)_{cm}}{f_{N,p} \big(\tilde\rho(S)\big)_{0m}} \\
&= f_{N,p} \left( \sum_m \frac{\tilde\rho(S)_{a,m} \tilde\rho(S)_{b,m} \tilde\rho(S)^{-1}_{c,m}}{ \tilde\rho(S)_{0,m}} \right)\\
&= f_{N,p} \big( {}_0{\tilde N_{ab}}^{~~~c} \big) 
= {}_0{\tilde N_{ab}}^{~~~c} .
\end{split}\end{align}
The fields in the Hecke image are aligned in the same way as before.
The first row of $f_{N,p} \big(\tilde\rho(\gamma)\big)$ still corresponds to the vacuum, in the sense of $\mathsf{T}_p \chi$.
We therefore confirm that the fusion rules are preserved under suitable Hecke operators, i.e. $\mathsf{T}_p$ with ${\bar p}^2 \equiv \ell^2 \pmod{N}$.
In summary,
\begin{align}\label{FusionRules:same}
{}_0{N_{ab}}^{c \, (p)} = {}_0{\tilde N_{ab}}^{~~~c} = {}_0{N_{ab}}^c .
\end{align}
This result will prove essential in establishing the polynomial equations for various RCFTs.

\subsection{$\mathsf{M}(3,5)$ as an example}
\label{subsec:M35}

We now illustrate the technique of effective picture using the minimal model $\mathsf{M}(3,5)$ as an example.
The non-unitary minimal model $\mathsf{M}(3,5)$ has (effective) central charge 
\begin{align}
c \big[\mathsf{M}(3,5)\big] = -3/5, \qquad
c_{\text{eff}} \big[\mathsf{M}(3,5)\big] = 3/5.
\end{align}
The primary fields are
\begin{align}
\phi_0, \, \phi_{-\frac{1}{20}}, \,\phi_{\frac{1}{5}}, \,\phi_{\frac{3}{4}},
\end{align}
where the subscripts denote the conformal weights.
$\phi_{-\frac{1}{20}}$ has the smallest conformal weight and is recognized as the minimal primary $\mathit{o}$. The vacuum field $\phi_0$ is the trivial simple current, while
$\phi_{\frac{3}{4}}$ has order 2 and permutes the primaries by
\begin{align}
\Big(\phi_0, \, \phi_{-\frac{1}{20}}, \,\phi_{\frac{1}{5}}, \,\phi_{\frac{3}{4}}\Big)
\times \phi_{\frac{3}{4}} =
\Big(\phi_{\frac{3}{4}},\,\phi_{\frac{1}{5}}, \, \phi_{-\frac{1}{20}},  \,\phi_0\Big).
\end{align}

The modular representation is
\beqs
\begin{align}
\rho^{\mathsf{M}(3,5)}(T)=&~
{\rm diag}\big( \xi_{40}, \xi_{40}^{-1}, \xi_{40}^{9}, \xi_{40}^{-9} \big) ,\\
\rho^{\mathsf{M}(3,5)}(S)=&
\begin{pmatrix}
x & y & -y & -x \\
y & x & x & y \\
-y & x & -x & y \\
-x & y & y & -x
\end{pmatrix} ,
\end{align}
\eeqs
where
\begin{align}
x=\sqrt{\frac{2}{5}} \sin\Big(\frac{2\pi}{5}\Big),\qquad
y=\sqrt{\frac{2}{5}} \sin\Big(\frac{\pi}{5}\Big).
\end{align}
With $N=40$, the quadratic subgroup is 
$\big[(\mathbb{Z}/N\mathbb{Z})^{\times}\big]^2
=\big\{1,9\big\}$.

With the trivial simple current, the GS equation does not hold for any 
$\ell^2 \in \big[(\mathbb{Z}/N\mathbb{Z})^{\times}\big]^2$.
The GS property requires the simple current $\phi_{\frac{3}{4}}$ and the quadratic element $\ell^2 \equiv 9 \pmod{N}$.
Following eq\eqref{rho:tilde}, one converts $\rho^{\mathsf{M}(3,5)}(\gamma)$ to the representation in the effective picture.
\beqs
\begin{align}
\rho^{\mathsf{\tilde M}(3,5)}(T)=&~
{\rm diag}\big( \xi_{40}^{-1}, \xi_{40}, \xi_{40}^{-9}, \xi_{40}^{9} \big) ,\\
\rho^{\mathsf{\tilde M}(3,5)}(S)=&
\begin{pmatrix}
x & -y & -y & x \\
-y & x & -x & y \\
-y & -x & -x & -y \\
x & y & -y & -x
\end{pmatrix} .
\end{align}
\eeqs
The effective twists are computed from $\rho^{\mathsf{\tilde M}(3,5)}(T)$ as
\begin{align}\label{M35twist:eff}
\tilde\theta =
1,\,\me\Big(\frac{1}{20}\Big),\,\me\Big(\frac{4}{5}\Big),\,\me\Big(\frac{1}{4}\Big).
\end{align}

In terms of field content, $\mathsf{\tilde M}(3,5)$ is viewed as the tensor product of affine algebra $(A_1,1)$ and $\overline{\mathsf{\tilde M}(2,5)}$. The latter is the complex conjugate of $\mathsf{\tilde M}(2,5)$ (Yang-Lee model), which has 
\begin{align}
c \big[\mathsf{\tilde M}(2,5)\big] = c_{\text{eff}} \big[\mathsf{M}(2,5)\big] = 2/5. 
\end{align}
As a result, the modular representation of $\mathsf{\tilde M}(3,5)$ is simply the Kronecker product of $(A_1,1)$ and $\overline{\mathsf{\tilde M}(2,5)}$. The $(A_1,1)$ MTC has the vacuum and the semion as primary fields. 
The semion has conformal weight $1/4$ and serves as the simple current in the tensor product structure.
Note that $\mathsf{M}(3,5)$ has conductor $N=40$.

The characters of $\mathsf{M}(3,5)$ are given by
\beqs
\begin{align}
\chi^{\mathsf{M}(3,5)}_0 (\tau) &= q^{\frac{1}{40}}
(1+ q^2+ q^3+ 2q^4+ 2q^5+\cdots) ,\\
\chi^{\mathsf{M}(3,5)}_{-\frac{1}{20}} (\tau) &= q^{-\frac{1}{40}}
(1+ q+ q^2+ 2q^3+ 3q^4+ 4q^5+\cdots) ,\\
\chi^{\mathsf{M}(3,5)}_{\frac{1}{5}} (\tau) &= q^{\frac{9}{40}}
(1+ q+ 2q^2+ 2q^3+ 3q^4+ 4q^5+\cdots) ,\\
\chi^{\mathsf{M}(3,5)}_{\frac{3}{4}} (\tau) &= q^{\frac{31}{40}}
(1+ q+ q^2+ 2q^3+ 2q^4+ 3q^5+\cdots).
\end{align}
\eeqs
Among the components, $\chi^{\mathsf{M}(3,5)}_{-\frac{1}{20}}$ contains the most singular term and corresponds to the minimal primary $\mathit{o}$, while $\chi^{\mathsf{M}(3,5)}_0$ is the vacuum character due to its leading term $q^{-c[\mathsf{M}(3,5)]/24}$.
Moreover, the true vacuum is invariant under the Poincar{\'e} group and in particular under translations. Hence, the Virasoro generator $L_{-1}$ annihilates the vacuum, i.e. $L_{-1} |0\rangle = 0$ and there is thus no $q^1$ term in the vacuum character.

The characters of $(A_1,1)$ and $\mathsf{M}(2,5)$ have Hecke images which were computed in  \cite{Hecke:2018}.
It is interesting to explore Hecke images of the $\mathsf{M}(3,5)$ characters as well.
Explicit computation by eq\eqref{Gp:formula} provides the list of $G^{\mathsf{M}(3,5)}_{\bar p}=\rho^{\mathsf{M}(3,5)}(\sigma_p)$ for all $p \in (\mathbb{Z}/N\mathbb{Z})^{\times}$. \\
When $p=1,11,29,39$,
\begin{align}
\rho^{\mathsf{M}(3,5)}(\sigma_p)=\mathbb{I}_4  ~,
\qquad p^2=1 \pmod{40}.
\end{align}
When $p=9,19,21,31$,
\begin{align}
\rho^{\mathsf{M}(3,5)}(\sigma_p)= - \mathbb{I}_4  ~,
\qquad p^2=1 \pmod{40}.
\end{align}
When $p=3,7,33,37$,
\begin{align}
\rho^{\mathsf{M}(3,5)}(\sigma_p)= 
\begin{pmatrix}
 0 & 0 & 1 & 0 \\
 0 & 0 & 0 & -1 \\
 -1 & 0 & 0 & 0 \\
 0 & 1 & 0 & 0 \\
\end{pmatrix},
\qquad p^2=9  \pmod{40}.
\end{align}
When $p=13, 17, 23, 27$,
\begin{align}
\rho^{\mathsf{M}(3,5)}(\sigma_p) =  -
\begin{pmatrix}
 0 & 0 & 1 & 0 \\
 0 & 0 & 0 & -1 \\
 -1 & 0 & 0 & 0 \\
 0 & 1 & 0 & 0 \\
\end{pmatrix},
\qquad p^2=9  \pmod{40}.
\end{align}


The four distinct values of $\rho^{\mathsf{M}(3,5)}(\sigma_p)$ form the cyclic group $C_4$ under multiplication. As we shall see later, $(A_1,3)$ is the Hecke image theory of $\mathsf{M}(3,5)$ under $\mathsf{T}_3$. The matrices $\rho^{(A_1,3)}(\sigma_p)$ are the same as $\rho^{\mathsf{M}(3,5)}(\sigma_p)$ with proper ordering of the basis. The authors in \cite{Coste:1993af} computed the four distinct values of $\rho^{(A_1,3)}(\sigma_p)$ and regarded $C_4$ as the Galois group on primary fields. 
But the Galois group of fusion rules we refer to is basically the permutation within the matrices $\{N_i | i \in \CI \}$, where the permutation is given by eq(19) in \cite{Coste:1993af} and $N_i$ is defined by eq\eqref{fusion_rule:matrix}.
Since $\rho(\sigma_p)$ only tells how the primary fields are shuffled given its non-zero entries, the overall sign of $\rho(\sigma_p)$ does not affect the field permutation. Hence the fusion rule automorphism is characterized by $\pm \rho^{\mathsf{M}(3,5)}(\sigma_p)$ in $\mathsf{M}(3,5)$ or $(A_1,3)$, and the Galois group of fusion rules is exactly $\CG=C_2$, in agreement with the group of quadratic residues $\big[(\mathbb{Z}/N\mathbb{Z})^{\times}\big]^2$.

Not every Hecke image corresponds to a unitary RCFT.
A unitary RCFT or MTC requires non-negative integral fusion coefficients that are determined by the Verlinde formula eq\eqref{Verlinde_formula}. 
If the constraint of unitarity is relaxed, there could be negative fusion coefficients though with positive $q$-series.
For simplicity, here we focus on the Hecke images which have interpretations as the characters of unitary RCFTs. They correspond to the series $p=3,7,33,37 \pmod{40}$.
The Hecke images $\mathsf{T}_p \chi^{\mathsf{M}(3,5)}$ with $p=3,7$ provide the characters of two affine Lie algebras.
\begin{align}\label{A1_3}
&\mathsf{T}_3 \chi^{\mathsf{M}(3,5)}=\chi^{(A_1,3)}, \\
\chi^{(A_1,3)}_0 &= q^{-\frac{3}{40}}
(1+ 3q+ 9q^2+ 22q^3+ 42q^4+ 81q^5+\cdots) ,
\tag{\ref{A1_3}a} \\
\chi^{(A_1,3)}_{\frac{3}{20}} &= q^{\frac{3}{40}}
(2+ 6q +18q^2+ 36q^3+ 78q^4+ 144q^5+\cdots) ,
\tag{\ref{A1_3}b} \\
\chi^{(A_1,3)}_{\frac{2}{5}} &= q^{\frac{13}{40}}
(3+ 9q+ 20q^2+ 45q^3+ 90q^4+ 170q^5+\cdots) ,
\tag{\ref{A1_3}c} \\
\chi^{(A_1,3)}_{\frac{3}{4}} &= q^{\frac{27}{40}}
(4+ 6q+ 18q^2+ 34q^3+ 72q^4+ 126q^5+\cdots) ;
\tag{\ref{A1_3}d} 
\end{align}
\begin{align}\label{C3_1}
&\mathsf{T}_7 \chi^{\mathsf{M}(3,5)}=\chi^{(C_3,1)}, \\
\chi^{(C_3,1)}_0 &= q^{-\frac{7}{40}}
(1+ 21q+ 126q^2+ 511q^3+ 1743q^4+\cdots) ,
\tag{\ref{C3_1}a} \\
\chi^{(C_3,1)}_{\frac{7}{20}} &= q^{\frac{7}{40}}
(6+ 70q+ 336q^2+ 1302q^3+ 4186q^4+\cdots) ,
\tag{\ref{C3_1}b} \\
\chi^{(C_3,1)}_{\frac{3}{5}} &= q^{\frac{17}{40}}
(14+ 105q+ 483q^2+ 1764q^3+ 5523q^4+\cdots) ,
\tag{\ref{C3_1}c} \\
\chi^{(C_3,1)}_{\frac{3}{4}} &= q^{\frac{23}{40}}
(14+ 78q+ 378q^2+ 1288q^3+ 4032q^4+\cdots) .
\tag{\ref{C3_1}d} 
\end{align}
The affine Lie algebras $(A_1,3)$ and $(C_3,1)$ have central charges $p\, c_{\text{eff}}$ with $p=3$ and $7$, respectively.
However there is no obvious way to realize the RCFTs for $p=33,37$, though the derived MTCs by Galois conjugation are unitary. 
They are perhaps intermediate vertex subalgebras similar to the $E_{7\frac{1}{2}}$ theory \cite{kawasetsu,sextonion}.
As expected, all four MTCs in this series enter into the classification of topological orders \cite{Wen:2015,Schoutens:2016}.
Notably Hecke images of $\chi^{\mathsf{M}(3,5)}$ can be solved from the MLDE eq\eqref{MLDE:order4}. 
The modular representation for $\mathsf{T}_p \chi^{\mathsf{M}(3,5)}$ is
\begin{align}
\rho^{(p)}(\gamma) =
f_{N,p}\left(
\rho^{\mathsf{\tilde M}(3,5)} (\gamma) \right).
\end{align}
In particular for $p=3,7,33,37$,
\begin{align}
\rho^{(p)}(S) =
\begin{pmatrix}
y & x & x & y \\
x & y & -y & -x \\
x & -y & -y & x \\
y & -x & x & -y
\end{pmatrix}
\end{align}
meets all the requirements of unitary MTC, i.e. the non-negative fusion coefficients and the quantum dimensions $d_i^{(p)} \geq 1$.

We infer the fusion rules in the Hecke images of $\mathsf{M}(3,5)$ by the analysis earlier.
They are expressed in terms of the matrices $N_i$ defined in eq\eqref{fusion_rule:matrix}, where a super-index labels the RCFT and a sub-index indicates the conformal weight as usual. The primary fields are arrayed in the same order as before.
\begin{align}
N_0^{\mathsf{M}(3,5)} 
= N_0^{\mathsf{\tilde M}(3,5)} 
= N_0^{(A_1,3)} 
= N_0^{(C_3,1)} 
= \II_4.
\end{align}
\begin{align}
N_{-\frac{1}{20}}^{\mathsf{M}(3,5)} = N_{\frac{1}{20}}^{\mathsf{\tilde M}(3,5)} = N_{\frac{3}{20}}^{(A_1,3)} = N_{\frac{7}{20}}^{(C_3,1)} 
= \left(
\begin{array}{cccc}
 0 & 1 & 0 & 0 \\
 1 & 0 & 1 & 0 \\
 0 & 1 & 0 & 1 \\
 0 & 0 & 1 & 0 \\
\end{array}
\right).
\end{align}
\begin{align}
N_{\frac{1}{5}}^{\mathsf{M}(3,5)} = N_{\frac{4}{5}}^{\mathsf{\tilde M}(3,5)} = N_{\frac{2}{5}}^{(A_1,3)} = N_{\frac{3}{5}}^{(C_3,1)} 
= \left(
\begin{array}{cccc}
 0 & 0 & 1 & 0 \\
 0 & 1 & 0 & 1 \\
 1 & 0 & 1 & 0 \\
 0 & 1 & 0 & 0 \\
\end{array}
\right).
\end{align}
\begin{align}
N_{\frac{3}{4}}^{\mathsf{M}(3,5)} = N_{\frac{1}{4}}^{\mathsf{\tilde M}(3,5)} = N_{\frac{3}{4}}^{(A_1,3)} = N_{\frac{3}{4}}^{(C_3,1)} 
= \left(
\begin{array}{cccc}
 0 & 0 & 0 & 1 \\
 0 & 0 & 1 & 0 \\
 0 & 1 & 0 & 0 \\
 1 & 0 & 0 & 0 \\
\end{array}
\right).
\end{align}
They agree perfectly with the fusion rules calculated from the modular data in these RCFTs.

Let us not forget that $\mathsf{M}(3,5)$, as a member of non-unitary minimal models, has long been known to describe critical phases of 2D classical statistical mechanics models, such as the ``restricted solid-on-solid'' (RSOS) models \cite{RSOS}. In condensed matter physics, $\mathsf{M}(3,5)$ is of particular interest as it describes the critical behavior of a chain of antiferromagnetically coupled Yang-Lee anyons \cite{Ardonne}. The earlier discovered Galois conjugation relations between Yang-Lee and Fibonacci anyons serve as a first example of the broader Galois symmetries induced by Hecke relations between different RCFTs we present in this paper.

\section{Duality Transformation of Conformal Blocks}
\label{sec:duality}
Besides modular invariance, duality is another distinctive property of RCFT. 
In this section, we describe the duality transformations in RCFT and build the formalism for probing Galois symmetry.
This section is largely a review of the literature, and set up the notation for the following section.

\subsection{Chiral vertex operators and conformal blocks}
\label{subsec:CVO}
In preparation for our discussion of duality, we first define the chiral vertex operators (CVOs). Their correlation functions are conformal blocks for physical correlation functions. The exchange symmetries of conformal blocks are described by duality transformations. See \cite{Moore:polynomial,Moore:naturality,Moore:classical} for mathematical details.

The physical Hilbert space $\CH_{\rm phys}$ is a direct sum over irreducible representations of $\CA\times \overline \CA$, as is reviewed in eq\eqref{H:decomp}. Every state in the decomposition transforms as the representation $\big( V_i, \overline{V_{\bar i}} \big)$.
The CVO is the intertwining operator for chiral representations, with dependence on the coordinate $z$ on the complex plane. Given three representations labeled by $i,j,k \in \CI$, 
we define the operator 
\begin{align}\label{chiral_operator_map}
\Phi_t(z) : (V_i)^{\vee}\otimes V_j \otimes V_k \rightarrow \mathbb{C},
\end{align}
where $ (V_i)^{\vee}$ is the dual of $V_i$. The representations are ordered such that $j,k$ refer to the incoming states and $i$ labels the outgoing one. Such operators are called of type $(i;j,k)$, and the subscript $t$ distinguishes between different operators of the same type.
In general the CVOs of type $(i;j,k)$ span a vector space $V^i_{jk}$, which has dimensionality
\begin{align}
\text{dim}\, V^i_{jk} = N_{jk i^{\vee}}={N_{jk}}^i .
\end{align}
The numbers ${N_{jk}}^i$ are the fusion rules determined by the Verlinde formula eq\eqref{Verlinde_formula}, and their dependence on the vacuum 0 is omitted occasionally.
The case ${N_{jk}}^i \leq 1$ contains most essential features of RCFT and affords a simpler description.
In this situation, there is only one operator of type $(i;j,k)$, which can be written as $\Phi^i_{jk}$ for brevity.

\begin{figure}[h]
		\centering
 \includegraphics[width=1.2in]{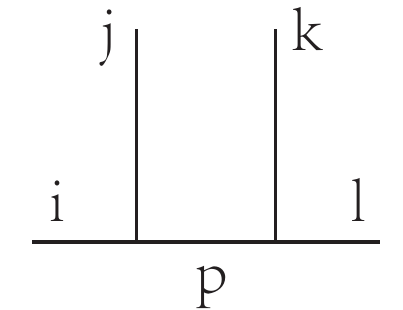}
        \caption{A geometric illustration of fusion, as the composition of two 4-point functions.}
            \label{fig:ConformalBlock}
\end{figure}

In RCFT  conformal blocks form a basis for physical 4-point functions. Each conformal block is computed by gluing two CVOs at points which we label as $z_2, z_3$, with the initial and the final state at $0$ and $\infty$ respectively \cite{Moore:1989vd}.
\begin{align}\label{ConformalBlock:ijkl}
\CF_p^{ijkl} (z_2, z_3) :=
\big\langle i \big| \Phi_{ip}^j(z_2) \Phi_{pl}^k(z_3) \big| l \big\rangle 
\end{align}
Figure \ref{fig:ConformalBlock} gives a graphical description, where the indices $i,j,k,l$ stand for the external legs while $p$ labels the field in the mediated channel. In the diagonal theory, the physical correlation function is
\begin{align}\label{phys_corre}
\big\langle \phi_i(\infty,\infty) \phi_j(z_2,\bar z_2) \phi_k(z_3,\bar z_3) \phi_l(0,0) \big\rangle 
= \sum_{p \in \CI} \mathsf{D}_p \, \big| \CF_p^{ijkl} (z_2, z_3) \big|^2,
\end{align}
where $\mathsf{D}_p$ are constants independent of $z$ and $\bar z$.

\subsection{Fusion and braiding symmetries}
\label{subsec:F_B_sym}
The axiom of duality states that physical correlation functions do not depend on the choice of the basis of conformal blocks. 
The conformal block for any diagram is a linear combination of conformal blocks for any other \cite{Moore:classical}. 
In particular, duality of the 4-point functions implies the existence of fusion and braiding matrices, which are induced by F- and B-moves respectively. 
When acting on $\CF_p^{ijkl} (z_2, z_3) $, the F- and B-moves cause the change
\begin{align}\label{F:map}
F
\begin{bmatrix}
j &k \\ i &l
\end{bmatrix} :&
\oplus_p V^i_{jp}\otimes V^p_{kl} \rightarrow \oplus_q V^i_{ql}\otimes V^q_{jk} ~,\\
\label{B:map}
B
\begin{bmatrix}
j &k \\ i &l
\end{bmatrix} :&
\oplus_p V^i_{jp}\otimes V^p_{kl} \rightarrow \oplus_q V^i_{kq}\otimes V^q_{jl} ~,
\end{align}
where the matrix elements $F_{pq}, B_{pq}$ specify the initial and the final terms in the direct sum \cite{Moore:classical}.
Any duality transformation are expressible by these two basic moves.
We will elucidate the fusion and the braiding matrix explicitly in terms of operator product expansion (OPE).

\begin{figure}[h]
		\centering
 \includegraphics[width=3.6in]{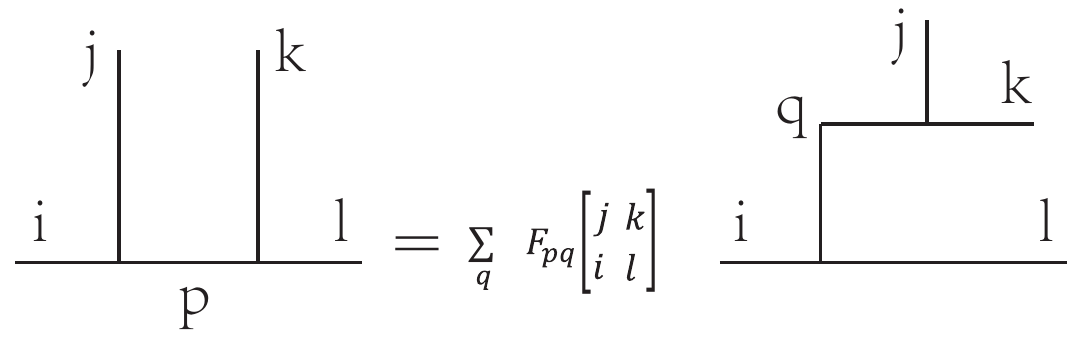}
        \caption{Fusion matrix between blocks. The labels of the matrix entries, i.e. $p$ and $q$, take the positions of the ``propagator".}
            \label{fig:FusionMatrix}
\end{figure}

Let $z_{ij}$ be shorthand for $z_i-z_j$.
The fusion matrix $F$ is defined by 
\begin{align}\label{F:def}
\Phi_{ip}^j(z_2) \Phi_{pl}^k(z_3)=
\sum_{q \in \CI} F_{pq}
\begin{bmatrix}
j &k \\ i &l
\end{bmatrix}
\sum_{Q\in V_q} \Phi^{q,Q}_{il}(z_3) \big\langle Q \big | \Phi^j_{qk}(z_{23}) \big | k \big\rangle ~,
\end{align}
where $Q\in V_q$ denotes the descendant states in the module $V_q$ \cite{Moore:1989vd}. To obtain the OPE on the right hand side, we use the translation and scaling invariance. 
Figure \ref{fig:FusionMatrix} characterizes the $s$-$t$ duality schematically.
Two successive F-moves are equivalent to the identity transformation, leading to the quadratic relation
\begin{align}\label{F:quadratic}
\sum_{q} 
F_{pq}
\begin{bmatrix}
j &k \\ i &l
\end{bmatrix} \,
F_{qp'}
\begin{bmatrix}
l &k \\ i &j 
\end{bmatrix}
=\delta_{pp'}~.
\end{align}

\begin{figure}[h]
		\centering
 \includegraphics[width=3.6in]{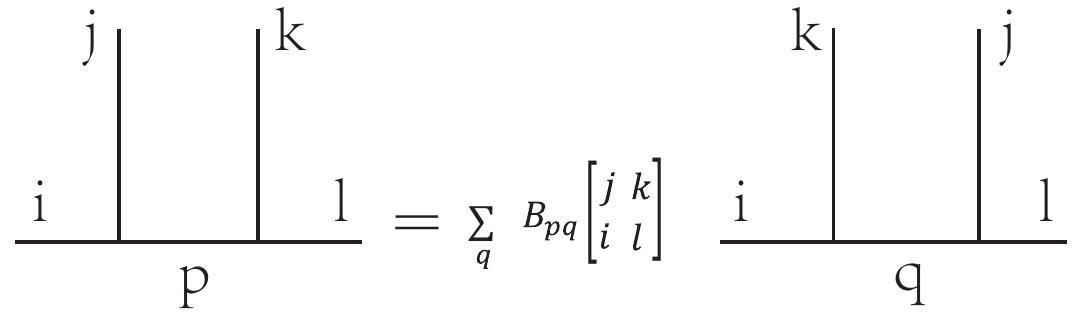}
        \caption{Braiding matrix between blocks. The labeling of the matrix entries, i.e. $p$ and $q$, take the positions of the ``propagator".}
            \label{fig:BraidingMatrix}
\end{figure}

The braiding matrix $B$ is defined by 
\begin{align}
\Phi_{ip}^j(z_2) \Phi_{pl}^k(z_3)=
\sum_{q \in \CI} B_{pq}
\begin{bmatrix}
j &k \\ i &l
\end{bmatrix}
\Phi_{iq}^k(z_3) \Phi_{ql}^j(z_2)
\end{align}
\cite{Moore:1989vd}. Figure \ref{fig:BraidingMatrix} provides the graphical illustration for the $s$-$u$ duality. 
In fact $B_{pq}$ is the monodromy matrix for the vector of blocks 
$\CF_p^{ijkl} (z_2, z_3)$ when $z_2$ circles around $z_3$.
The braiding matrix is independent of $z$ in each connected region of the common domain. Given two regions separated by a branch cut, there are two transformations 
\begin{align}
B(\epsilon), \quad
\epsilon= \text{sgn}\big(\Im (z_{23}) \big)
\end{align}
with the consistency condition
\begin{align}\label{B:quadratic}
\sum_{q} 
B_{pq}
\begin{bmatrix}
j &k \\ i &l
\end{bmatrix} (\epsilon) \,
B_{qp'}
\begin{bmatrix}
k & j \\ i & l 
\end{bmatrix} (-\epsilon)
=\delta_{pp'}~.
\end{align}
It should be stressed that $B^2$ is not the identity matrix because of the cuts. If the sign $\epsilon$ is omitted, we are referring to $B(+)$.
For a coupling $t$ of type $(i;j,k)$,
we define the operators $\Omega$ and $\Theta$
\beqs
\begin{alignat}{2}
\Omega(\pm) &: V^i_{jk}\cong V^i_{kj}, \qquad
& \Omega(\pm)(t)=  \me^{\pm \mi\pi\Delta_t} \varsigma_{2 3}(t) ,\label{Omega}\\
\Theta(\pm) &: V^i_{jk}\cong V^{k^{\vee}}_{j i^{\vee}}, \qquad
& \Theta(\pm)(t)= \varsigma_{1 3}(\me^{\pm \mi\pi\Delta_t} t), \label{Theta}
\end{alignat}
\eeqs
Here $\Delta_t =\Delta_j+\Delta_k-\Delta_i$, with $\Delta_i$ the smallest $L_0$ eigenvalue of the states in $V_i$.
$\varsigma_{ij}$ is a transposition of $i$ and $j$ with $\varsigma_{ij}^2=1$.
The extra phase $\me^{\pm \mi\pi\Delta_t}$ compensates the phase arising from swapping the external legs, done by $z \rightarrow e^{\pm \mi \pi}z$ depending on the cut.
$\Omega$ and $\Theta$ are special cases of the B-move. 
The operation $\Omega$ is also referred to as the R-move, and its eigenvalues are called the braiding eigenvalues or the $R$-matrices.

\begin{figure}[h]
		\centering
 \includegraphics[width=3in]{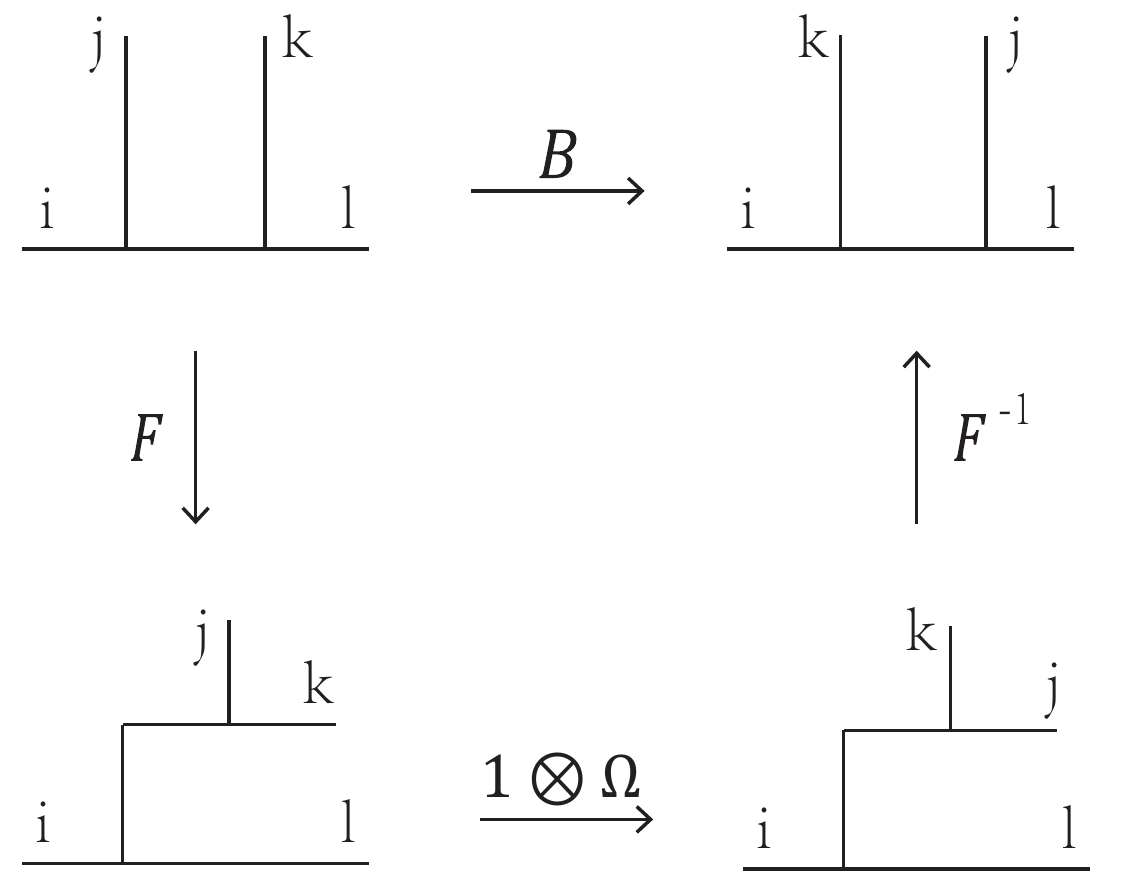}
        \caption{A simple loop transformation of conformal block.}
            \label{fig:Loop_FourPointFunction}
\end{figure}

We start with a 4-point function and perform the duality transformations of the CVOs in two ways as depicted in Figure \ref{fig:Loop_FourPointFunction}. 
Ending in the same configuration, we build
\begin{align}\label{B:FRF2}
B_{pp'}
\begin{bmatrix}
j &k \\ i &l
\end{bmatrix}(\epsilon) 
= \sum_{q \in \CI}   
F_{pq}
\begin{bmatrix}
j &k \\ i &l
\end{bmatrix} \,
\me^{-\mi\pi\epsilon(\Delta_k + \Delta_j - \Delta_q)} \,
F_{qp'}
\begin{bmatrix}
l & j \\ i & k
\end{bmatrix} ~,
\end{align}
or symbolically
\begin{align}\label{F_Omega_F}
B(\epsilon)= F^{-1} \big[1\otimes\Omega(-\epsilon)\big] F .
\end{align}
The B-move is simply a combined operation of F- and R-moves.
As a consequence, eigenvalues of the $B$-matrices are square roots of mutual locality factors and are deduced as half-monodromies.

The duality matrices are usually determined as follows.
We first compute the fusion rules by the Verlinde formula and find all the fusion channels. Given any five-point function, we can formulate different sequences of F-moves from the same starting fusion basis decomposition to the same ending decomposition.
These consistency conditions build the polynomial equations called the \textit{pentagon equations}.
The solution to the pentagon equations is organized into the $F$-matrices, whose entries are known as the $6j$ symbols \cite{Rowell:2009}.
Likewise consistency relations arise if the R-moves act on the fusion space of three particles in different ways, ending in the commutative hexagon diagrams. 
The hexagon diagrams contain both F- and R-moves, making the braidings compatible with the fusions. They give rise to the \textit{hexagon equations}.
In practice, we first solve the pentagon equations to gain all the fusion matrices. We then insert the solved fusion matrices into the hexagon equations and determine all the braiding eigenvalues.
Despite the several sets of solutions, we pick the desired one by inspecting typical braiding eigenvalues in that MTC. (A complete set of fusion matrices does not determine the MTC, and could incorporate many sets of consistent braiding eigenvalues.)

Using global conformal symmetry, we rewrite the correlator in terms of the cross-ratio
$z= z_{12}z_{34} / z_{13}z_{24}$. If the coordinates of the external legs are chosen to be
\begin{align}
z_1=\infty,\quad z_2=1, \quad z_3=z,\quad z_4=0 ,
\end{align}
the cross-ratio reduces to $z$. There are two other cross-ratios
\begin{align}
1-z=\frac{z_{14}z_{23}}{z_{13}z_{24}},\qquad
\frac{z}{1-z}=\frac{z_{12}z_{34}}{z_{14}z_{23}}~.
\end{align}
Duality transformations are done by permuting the positions of CVOs. The F-move results in the permutation $\varsigma_{1234}$ on the external legs, which amounts to $z \rightarrow 1-z$ on the coordinates.
The R-move is simply done by the transposition $\varsigma_{23}$, which takes $z$ to $1/z$.
Thus the B-move causes the transformation $z \rightarrow  z/(z-1)$, courtesy of eq\eqref{F_Omega_F}.

Without loss of generality, we consider the 4-point function 
\begin{align}\label{4_point:real}
\CG (z_i,\bar z_i) = \big\langle \phi_A(z_1,\bar z_1) \phi_A(z_2,\bar z_2) \phi_A(z_3,\bar z_3) \phi_A(z_4,\bar z_4) \big\rangle 
\end{align}
of a real primary field $\phi_A$. By conformal symmetry, $\CG (z_i,\bar z_i)$ factors into
\begin{align}\label{correlation:factor}
\CG (z_i,\bar z_i) = (z_{14}z_{32} \bar z_{14} \bar z_{32})^{-2h_A} \,G(z,\bar z),
\end{align}
where $h_A$ is the conformal weight of $\phi_A$. For convenience we adopt the shorthand notation
\begin{align}
f_{\alpha}(z)=\CF_{\alpha}^{AAAA}(z),
\end{align}
where $\CF_{\alpha}^{AAAA}(z)$ is perceived as the conformal block with the $z_{ij}$ powers factored out.
The conformally invariant part $G(z,\bar z)$ is a sum over conformal blocks $f_{\alpha}$:
\begin{align}\label{ConformalBlocks:sum}
G(z,\bar z)=\sum_{\alpha\in \CI} d_{AA\alpha}^2 \,f_{\alpha}(z) \bar f_{\alpha}(\bar z)
=\sum_{\alpha\in \CI} d_{AA\alpha}^2 \,\big| f_{\alpha}(z) \big|^2,
\end{align}
where $d_{AA\alpha}$ are the OPE coefficients. Unitary RCFTs require $d_{AA\alpha}^2$ to be positive, while $d_{AA\alpha}^2$ could be negative in a non-unitary RCFT.
The normalization of conformal blocks depends on the OPE coefficients, and only the product $\big| d_{AA\alpha} \,f_{\alpha}(z) \big|$ is definite. 
For this reason, we have freedom in choosing the off-diagonal entries of the $F$- and the $B$-matrices. 
Such freedom is referred to as a change of gauge \cite{Moore:1989vd}. The gauge transformation is parameterized by the relative fugacity matrix $\Lambda={\rm diag} \big( \lambda^2_{\alpha} \big)$, and takes the form 
\begin{align}\label{gauge_trans}
f_{\alpha}(z) \rightarrow \lambda_{\alpha}\, f_{\alpha}(z) ,
\qquad
F \rightarrow \Lambda^{-1}\, F \, \Lambda ,
\end{align}
where $F$ is any fusion matrix \cite{Freedman}. In the literature, the conventional gauge is chosen such that the $F$-matrices are symmetric.
Furthermore, whether or not an entry of the $F$-matrix vanishes is a gauge-invariant property \cite{Bonderson:thesis}.

To describe the gauge dependence, we take the Fibonacci-type fusion rule $\phi\times\phi=I+\phi$ as an example. The nontrivial fusion matrix reads
\begin{align}
F
\begin{bmatrix}
\phi & \phi \\ \phi & \phi
\end{bmatrix}
=
\begin{pmatrix}
a_{\pm} & 1 \\ a_{\pm} & -a_{\pm}
\end{pmatrix} ,
\end{align}
where $a_{\pm}=(-1\pm\sqrt{5})/2$ \cite{Moore:classical}. The choice of $a_+$ corresponds to the $G_2$ or the $F_4$ theory.
While the choice of $a_-$ yields an imaginary OPE coefficient, thus any RCFT with this monodromy is non-unitary. This verifies the non-unitarity of the Yang-Lee theory and the $E_{7\frac{1}{2}}$ theory.
If we choose the symmetric normalization, the $F$-matrix takes the familiar form as in \cite{Ardonne}.
\begin{align}
F_{\rm sym}
\begin{bmatrix}
\phi & \phi \\ \phi & \phi
\end{bmatrix}
=
\begin{pmatrix}
a_{\pm} & \sqrt{a_{\pm}} \\ \sqrt{a_{\pm}} & -a_{\pm}
\end{pmatrix} .
\end{align}
 
The conformal fields and the correlation functions are manifestly gauge invariant \cite{Moore:polynomial}.
It is gauge invariant as well for the pentagon and hexagon system of equations, i.e. the polynomial equations originating from various closed loop diagrams. For any solution to these equations, there exists a continuous family of solutions that are gauge equivalent to it.

\section{Duality Matrices and Galois Symmetry}
\label{sec:Fusion_and_Braiding}
Fusion and braiding are two basic duality transformations, as introduced in the last section.
In this section we demonstrate how the duality matrices are related in different Hecke image theories, whose MTCs are Galois conjugates.

\subsection{Fusion Matrices} 
\label{subsec:Fusion}
The fusion matrices inherit the Galois symmetry from the pentagon equations reviewed in the last section.
We begin the analysis by visiting the two-channel fusion, which affords explicit calculation of the conformal blocks. The fusion matrices computed thereof obey the Galois symmetry consistently. We then study the MTCs of some familiar RCFTs as evidence for general cases.
For simplicity we will confine ourselves to the fusion rules for which each fusion coefficient ${}_0{N_{ij}}^k$ equals 0 or 1. 

\subsubsection{Analytical results in two-channel fusion}
\label{subsubsec:analyic}
The physical correlation function eq\eqref{phys_corre} remains invariant under the crossing $z\rightarrow 1-z$.
Meanwhile, the holomorphic conformal blocks transform into themselves as
\begin{align}\label{1-z}
\CF_p^{ijkl}(1-z)=\sum_{q \in \CI} \CM_{pq} \, \CF_q^{iljk} (z).
\end{align}
The fusion matrix is computable, provided $ \CF_q^{ijkl} (z)$ is known.
It can be taken to a unitary matrix by gauge transformation for unitary RCFT, which amounts to choosing an orthonormal basis for the conformal blocks. Then the matrix elements $\CM_{pq}$ appear as the probability amplitudes.

We explain the idea with the 4-point function of a real primary $\phi_A$. 
Assume that there are at most two conformal blocks as is true for a number of RCFTs. The OPE of $\phi_A$ with itself must contain the identity operator, since $\phi_A$ is real and Hermitian. The assumed fusion rule would be
\begin{align}
\phi_A \times \phi_A = I +\phi_B,
\end{align}
where the identity $I$ and one other field $\phi_B$ flow in the intermediate channels. Denote by $h_A$ and $h_B$ the conformal weights of $\phi_A$ and $\phi_B$ respectively.
We shall calculate the conformal blocks of $\langle\phi_A\phi_A\phi_A\phi_A\rangle$ and extract the fusion matrices following the analytical approach in \cite{Mukhi:bootstrap}.

In order for $\langle\phi_A\phi_A\phi_A\phi_A\rangle$ to be non-vanishing, there are restrictions on the fusion channels. In particular, 
\begin{align}
\mathsf{N} = 8h_A+1-3h_B
\end{align}
must be a non-negative integer.
For a RCFT with finitely many chiral primaries, each primary field reorganizes an infinite number of Virasoro primaries.
Referring to the definition eq\eqref{F:def}, we thus need to consider the descendant states of the chiral primary in the intermediate channel.
For $0 \leq n \leq \mathsf{N}$, the integer $n$ labels the lowest secondary that flows in the $\phi_B$ channel, while $\mathsf{N} - n$ measures that in the vacuum channel. 
The part $G(z,\bar z)$ in the correlation function eq\eqref{correlation:factor} is expanded into the irreducible components $G^{(n)}(z,\bar z)$ labeled by $n$. With the given fusion rules, each $G^{(n)}$ is the sum of two conformal blocks
\begin{align}
G^{(n)}(z,\bar z)=\sum_{\alpha=0,1} 
\Big( d_{AA\alpha}^{(n)} \Big)^2 \,f^{(n)}_{\alpha}(z) \bar f^{(n)}_{\alpha}(\bar z) .
\end{align}
Here, the index $\alpha=0,1$ labels the vacuum component and the $\phi_B$ channel respectively.
For each $n$, the conformal block $f^{(n)}_{\alpha}(z)$ solves the differential equation  
\begin{align}\begin{split}
\label{DiffEq:f}
&\frac{\md^2}{\md z^2} f +\frac{2}{3} \Big(2h_A+1+\frac{\mathsf{N}}{2} \Big) \Big(\frac{1}{z}+\frac{1}{z-1}\Big) \frac{\md}{\md z} f 
+\left\{- \frac{2}{3}h_A (2h_A+1-\mathsf{N}) \Big(\frac{1}{z^2}+\frac{1}{(z-1)^2}\Big)\right. \\
&\left. +\Big(\frac{4h_A}{3}(2h_A+1-2n+\mathsf{N})+\frac{1}{3}(\mathsf{N}-n)(1+3n-\mathsf{N})\Big)\frac{1}{z(z-1)}\right\} f = 0 .
\end{split}\end{align}
This differential equation arises from studying the singular behavior of Wronskians, without knowledge of null vectors \cite{Mukhi:bootstrap}.
It is a variant of the hypergeometric equation and admits two fundamental solutions around the point $z=0$, i.e.
\beqs\label{DiffEq:solution}
\begin{align}
f^{(n)}_0(z)&=\big[z(1-z)\big]^{-2h_A}
\, _2F_1 (\mathfrak a, \mathfrak b; \mathfrak c; z), \\
f^{(n)}_1(z)&=\CN^{(n)} \big[z(1-z)\big]^{-2h_A}
z^{ 1 - \mathfrak c } \, _2F_1 (\mathfrak a - \mathfrak c + 1,\mathfrak b - \mathfrak c + 1; 2 - \mathfrak c; z),
\end{align}
\eeqs
where 
\begin{align}
\mathfrak a = \frac{1-4h_A-\mathsf{N}}{3}+n , \quad
\mathfrak b = -4h_A+\mathsf{N}-n , \quad
\mathfrak c = \frac{2(1-4h_A)+\mathsf{N}}{3}.
\end{align}
$_2F_1(\mathfrak a, \mathfrak b; \mathfrak c; z)$ is the hypergeometric function and $\CN^{(n)} $ is a normalization constant. 
The singularities in various coincident limits confirm that $f^{(n)}_0(z)$ and $f^{(n)}_1(z)$ correspond to the intermediate channels $I$ and $\phi_B$ respectively.

Under the crossing $z\rightarrow 1-z$, the conformal blocks transform via the fusion matrix $\CM^{(n)}$, which is just the transformation law of the hypergeometric function.
\begin{align}\def\arraystretch{1.3}
\begin{pmatrix}
f^{(n)}_0(1-z) \\f^{(n)}_1(1-z)
\end{pmatrix}
= \begin{pmatrix}
\CM^{(n)}_{00} &  \CM^{(n)}_{01}  \\
\CM^{(n)}_{10} &  \CM^{(n)}_{11}
\end{pmatrix}
\begin{pmatrix}
f^{(n)}_0(z) \\ f^{(n)}_1(z)
\end{pmatrix} 
\end{align}
Shifting $n$ by one unit flips the sign of $\CM^{(n)}$.
The case $n=\mathsf{N}$ is most relevant, where it is exactly the Virasoro vacuum that flows in the conformal primary of the identity.\footnote{We thank S. Mukhi for confirming this fact.}
The fusion matrix has the diagonal entries
\begin{align}\label{00,11}
\CM^{(n=\mathsf{N})}_{00}=-\CM^{(n=\mathsf{N})}_{11}=\frac{\sin\big[(h_B-4h_A)\pi\big]}{\sin\big[h_B \pi\big]},
\end{align}
which are gauge invariant.
Though the off-diagonal elements depend on the relative normalization $\CN^{(n)} $, one has the fixed product
\begin{align}
\CM^{(n=\mathsf{N})}_{10} \, \CM^{(n=\mathsf{N})}_{01}=
\frac{\sin\big[(2h_B-4h_A)\pi\big] \sin\big[4h_A\pi\big]}{\sin^2 \big[h_B \pi\big]} ,
\end{align}
which is obviously in $\IQ[\xi_N]$.
Appropriate choice of $\CN^{(n)}$ makes the fusion matrix $\CM^{(n)}$ unitary, yielding $\CM^{(n)}_{01} = \CM^{(n)}_{10}$. It corresponds to the symmetric normalization. Alternatively, it is possible to choose $\CM^{(n)}_{10}$ and $\CM^{(n)}_{01}$ such that they both sit in $\IQ[\xi_N]$.

The analysis of the fusion matrices applies to the effective picture and the Hecke images as well.
It is noteworthy that the conformal weights enter into the parameters of the conformal blocks.
Upon the Hecke operation $\mathsf{T}_p$, the effective conformal weights $\tilde h$'s get multiplied by $p$ modulo $\IZ$, namely
\begin{align}
h_A^{(p)} \equiv p\, \tilde h_A,\quad
h_B^{(p)} \equiv p\, \tilde h_B
\quad \pmod{1}.
\end{align}
Equivalently the twists are acted with the Frobenius map $f_{N,p}$.
In the Hecke image theory, the fusion matrix has the diagonal entries
\begin{align}
\frac{\sin\big[ (h_B^{(p)}-4h_A^{(p)})\pi \big]}{\sin\big[h_B^{(p)} \pi\big]}
=\frac{\sin\big[p(\tilde h_B-4\tilde h_A)\pi \big]}{\sin\big[p\,\tilde h_B \pi\big]}=
f_{N,p} \left( \frac{\sin\big[(\tilde h_B-4\tilde h_A)\pi\big]}{\sin\big[\tilde h_B \pi\big]} \right).
\end{align}
The effective conformal weights $\tilde h$ are evaluated by eq\eqref{h:tilde}.
While the selection rule demands that $2 Q_J(A)\equiv Q_J(B) + Q_J(0) \pmod{1}$.
These relations help to establish that
\begin{align}
 \frac{\sin\big[(\tilde h_B-4\tilde h_A)\pi \big]}{\sin\big[\tilde h_B \pi\big]}
 = f_{N,\ell^2} \left( \frac{\sin[( h_B - 4 h_A)\pi]}{\sin[ h_B \pi]} \right).
\end{align}
The Frobenius maps in the two steps combines to
\begin{align}
 f_{N,\ell^2} \circ  f_{N,p} = f_{N,\bar p} ~,
\end{align}
which is precisely the map between the modular representations upon the Hecke operation $\mathsf{T}_p$. 
It confirms that $f_{N,\bar p}$ transforms the diagonal entries of the fusion matrix.
The off-diagonal entries are not uniquely fixed. We could let them undergo the same Frobenius map if they are in $\IQ[\xi_N]$.
The determinant, as well as various polynomial equations of the $F$-matrix, are maintained under the Frobenius map.
By doing so, the normalizations of conformal blocks are naturally fixed in both the effective picture and the image theory.

Let us consider the general case of the fusion with $m$ channels ($m \geq 2$).
For RCFTs with multi-component primaries, their conformal blocks satisfy the BPZ equation \cite{BPZ}. The solutions of this equation are known to be hypergeometric functions. Therefore, we expect to extend what we have worked out to these cases as well. However, RCFTs with $m$ fusion channels, as appeared in theories such as WZW theories with high levels or latter members of minimal model series, come with at least $m$ types of anyons. This makes their physical realizations hard to achieve. The cases of complex primaries can be worked out similarly \cite{Mukhi:bootstrap}. We will leave them for future work.  


\subsubsection{Galois symmetry in fusion matrices}
\label{subsubsec:Galois_fusion}
We mentioned the philosophy of solving the $F$- and the $B$-matrices in Section \ref{sec:duality}.
For a given set of fusion rules, the solutions are discrete with the fixed gauge, including both unitary and non-unitary RCFTs.
For instance, there are eight distinct solutions with the Ising-type fusion rules \cite{Kitaev:anyon,Schoutens:2016}.
All of them can be realized by affine $\text{spin}(p)$ at level 1 with $p\in \big(\IZ/16\IZ)^{\times}$, thereby being unitary.
However, the Ising model has conductor $N=48$, and there exist Hecke images $\mathsf{T}_p \chi^{\rm Ising}$ for all prime $p$ with $\text{gcd}(p,48)=1$.
For all $p<24$ the Hecke images are the characters of  affine $\text{spin}(p)$ algebras at level 1 \cite{Hecke:2018}. 
While $\mathsf{T}_p \chi^{\rm Ising}$ is a shift from the $\text{spin}(p)$ characters for $24<p<48$:
\begin{align}
\begin{pmatrix}
\chi_0 \\ \chi_{v} \\ \chi_{s}
\end{pmatrix}^{{\rm spin}(p)}
-
\begin{pmatrix}
\chi_0 \\ \chi_{v} \\ \chi_{s}
\end{pmatrix}^{Y=(p)}
=p \,\CP \cdot
\begin{pmatrix}
\chi_0 \\ \chi_{v} \\ \chi_{s}
\end{pmatrix}^{{\rm spin}(p-24)}   ,
\end{align}
where
\begin{align}
\CP=\left(
\begin{array}{ccc}
 0 & 1 & 0 \\
 1 & 0 & 0 \\
 0 & 0 & -1 \\
\end{array}
\right)~.
\end{align}
The super-index $Y=(p)$ represents the Hecke image under $\mathsf{T}_p$. The sub-indices $0,v,s$ of $\chi$ stand for the vacuum, vector and spinor representation respectively.
The VOA of $\mathsf{T}_p \chi^{\rm Ising}$ sits in the $\text{spin}(p)$ MTC and therefore obeys the same duality transformations as the $\text{spin}(p)$ theory.

In the RCFT with character $\mathsf{T}_p \chi^{\rm Ising}$, we consider the correlation function $\big\langle \sigma^{(p)}\sigma^{(p)}\sigma^{(p)}\sigma^{(p)} \big\rangle$, where $\sigma^{(p)}$ is the spin field.
Denote the vacuum by $I^{(p)} $ and the fermion field by $\psi^{(p)}$.
The fusion rules are isomorphic to those of the Ising model, in particular 
\begin{align}
\sigma^{(p)} \times \sigma^{(p)} = I^{(p)} + \psi^{(p)}.
\end{align}
Hence, there are two conformal blocks with $I^{(p)}$ and $\psi^{(p)}$ as the intermediate channels. 
The associated fusion matrix is evaluated from the analytic method. A distinguished entry is
\begin{align}
F_{00} = 
\frac{\sin \big[(h_B^{(p)}-4h_A^{(p)})\pi \big]}{\sin \big[h_B^{(p)} \pi \big]}
= \cos \Big( \frac{p}{4} \pi \Big)
=\Big(\frac{2}{p}\Big) \frac{1}{\sqrt{2}} \, ,
\end{align}
where $\big( \frac{a}{n} \big)$ stands for the Jacobi symbol.
The entire fusion matrix reads
\begin{align}\label{fusion_sigma}
F
\begin{bmatrix}
\sigma^{(p)} & \sigma^{(p)} \\ \sigma^{(p)} & \sigma^{(p)} 
\end{bmatrix}
= \Big(\frac{2}{p}\Big)\cdot
\frac{1}{\sqrt{2}}
\begin{pmatrix}
1 & 1 \\ 1 & -1
\end{pmatrix} \,.
\end{align}
Similar results are listed in \cite{Kitaev:anyon}. With the property
\begin{align}
f_{48,\bar p} (\sqrt{2}) = \Big(\frac{2}{\bar p}\Big) \sqrt{2}
= \Big(\frac{2}{p}\Big) \sqrt{2}~,
\end{align}
eq\eqref{fusion_sigma} is translated to 
\begin{align}
F
\begin{bmatrix}
\sigma^{(p)} & \sigma^{(p)} \\ \sigma^{(p)} & \sigma^{(p)} 
\end{bmatrix}
= f_{48,\bar p} \left(
F
\begin{bmatrix}
\sigma & \sigma \\ \sigma & \sigma 
\end{bmatrix}
\right)
=\Big(\frac{2}{p}\Big) \, F
\begin{bmatrix}
\sigma & \sigma \\ \sigma & \sigma 
\end{bmatrix}.
\end{align}
The parity $\Big(\frac{2}{p}\Big) = \pm 1$ is critical and cannot be gauged away. 
In the MTC perspective, this sign corresponds to the Frobenius-Schur indicator (FSI). In general,
\begin{align}\label{FSI_d}
F_A \equiv F_{00}
\begin{bmatrix}
A & A \\ A & A
\end{bmatrix}= 1/\kappa_A d_A ,
\end{align}
where $d_A$ and $\kappa_A$ are the quantum dimension and the FSI of the primary field $\phi_A$ respectively \cite{Moore:naturality,Moore:polynomial}.
We study the FSI in more detail later.

There is a mathematical explanation for the above example. As shown in Section \ref{subsec:FusionRules}, the Hecke image theories have identical fusion rules, therefore the duality matrices obey the same set of polynomial equations.
The Galois symmetry of fusion matrices originates from the algebraic structure in pentagon equations.
By Ocneanu rigidity \cite{Ocneanu,Etingof:2002}, for any set of fusion rules there are only finitely many gauge equivalence classes of solutions to the polynomial equations.
We have a finite number of solutions with the fixed gauge \cite{Bonderson:thesis}. For the pentagon equations, each solution corresponds to an individual MTC and is characterized by $\left\{\sqrt{d_i} \right\}$, where $d_i$ are the quantum dimensions in that MTC.
Since a certain finite extension of $\IQ$ governs all the solutions \cite{Rowell:2009}, any solution is believed to have Galois-conjugate partners which correspond to other MTC solutions, and therefore $d_i$ in different solutions are related by Galois conjugation.
In retrospect, Galois conjugations of any solution respect the algebraic equations and the structure of MTC. Therefore, the same fusion rules hold, and the polynomial equations of the $F$-matrices are preserved.

We first examine the derived MTC from the Yang-Lee theory.
For the Fibonacci-type fusion rule $\phi\times\phi=I+\phi$, there are a total of four MTC solutions. They correspond to the Yang-Lee, $G_2$, $F_4$ and $E_{7\frac{1}{2}}$ theory respectively, with the common conductor $N=60$. The less-known $E_{7\frac{1}{2}}$ is an intermediate vertex subalgebra \cite{kawasetsu,sextonion}.
In each of the four MTCs, the entry $F_{00} = F_{\phi} $ is calculated by eq\eqref{00,11}. These entries are indeed related via the Frobenius maps $f_{N,p}$, explicitly
\begin{equation}\def\arraystretch{1.3}
\begin{array}{c@{\hspace{0.3cm}}|@{\hspace{0.3cm}}c@{\hspace{0.6cm}}c@{\hspace{0.6cm}}c@{\hspace{0.6cm}}c}
& {\rm YL}  & G_2 & F_4 & E_{7+\frac{1}{2}} \\
\hline
1/F_{\phi} & -1/g & g & g & -1/g \\ 
f_{N,p} & f_{60,1} & f_{60,7} & f_{60,13} & f_{60,19}
\end{array}
\quad ,
\end{equation}
where $g$ is the golden ratio.
\begin{align}
g=\me^{2\pi\mi/5}+\me^{-2\pi\mi/5}+1
=\me^{\pi\mi/5}+\me^{-\pi\mi/5}
=(1+\sqrt{5})/2 ~ .
\end{align}
Remarkably, the property $d_i \, d_j=\sum_k {}_0{N_{ij}}^k \, d_k$ implies the quadratic equation $x^2 = 1 + x$, which admits $g$ and $-1/g$ as Galois-conjugate solutions. 

\begin{sidewaystable}[!h]
\def\arraystretch{1.3}
\centering
\begin{tabular}{c||c|c|c|c}
\toprule
RCFT  & $\mathsf{M} (3,5)$ & $ \tilde{\mathsf{M}} (3,5)$  & $(A_1,3)$ & $(C_3,1)$ \\ 
\hline
unitarity & non-unitary &  ---   &  unitary &  unitary  
\\ 
$c $  &  $-3/5$ &  $3/5$   &  $9/5$ &  $21/5$
\\ 
field content 
&  $\phi_0, \phi_{\frac{-1}{20}}, \phi_{\frac{3}{4}}, \phi_{\frac{1}{5}}$
&  $\tilde\phi_0, \tilde\phi_{\frac{1}{20}}, \tilde\phi_{\frac{1}{4}}, \tilde\phi_{\frac{4}{5}}$   
&  $\varphi_0, \varphi_{\frac{3}{20}}, \varphi_{\frac{3}{4}}, \varphi_{\frac{2}{5}}$  
&  $\psi_0, \psi_{\frac{7}{20}}, \psi_{\frac{3}{4}}, \psi_{\frac{3}{5}}$  
\\ 
(anti-)semion orbit
& $\phi_{\frac{-1}{20}} \times \phi_{\frac{3}{4}} = \phi_{\frac{1}{5}}$
& $\tilde\phi_{\frac{1}{20}} \times \tilde\phi_{\frac{1}{4}} = \tilde\phi_{\frac{4}{5}}$
& $\varphi_{\frac{3}{20}} \times \varphi_{\frac{3}{4}} = \varphi_{\frac{2}{5}}$
& $\psi_{\frac{7}{20}} \times \psi_{\frac{3}{4}} = \psi_{\frac{3}{5}}$
\\
type I fusion
& $\phi_{\frac{-1}{20}} \times \phi_{\frac{-1}{20}}=\phi_0 + \phi_{\frac{1}{5}}$
& $\tilde\phi_{\frac{1}{20}} \times \tilde\phi_{\frac{1}{20}}=\tilde\phi_0 + \tilde\phi_{\frac{4}{5}}$   
& $\varphi_{\frac{3}{20}} \times \varphi_{\frac{3}{20}}=\varphi_0+\varphi_{\frac{2}{5}}$ 
& $\psi_{\frac{7}{20}} \times \psi_{\frac{7}{20}}=\psi_0+\psi_{\frac{3}{5}}$ 
\\ 
type II fusion  
& $\phi_{\frac{1}{5}} \times \phi_{\frac{1}{5}}=\phi_0 + \phi_{\frac{1}{5}}$ 
& $\tilde\phi_{\frac{4}{5}} \times \tilde\phi_{\frac{4}{5}}=\tilde\phi_0 + \tilde\phi_{\frac{4}{5}}$   
& $\varphi_{\frac{2}{5}} \times \varphi_{\frac{2}{5}}=\varphi_0+\varphi_{\frac{2}{5}}$ 
& $\psi_{\frac{3}{5}} \times \psi_{\frac{3}{5}}=\psi_0+\psi_{\frac{3}{5}}$ \\
\bottomrule
\end{tabular}
\caption{\label{tab:M35} This table summarizes some data for RCFTs derived from the $\mathsf{M}(3,5)$ MTC. As always, the subscripts of primary fields denote the conformal weights. 
The field content is also evaluated in the effective picture of $\mathsf{M}(3,5)$, so are the fusion rules. 
The approach of effective central charge is essentially the Galois conjugation plus an additional simple current permutation. 
The data in MTC, including $F$- and $B$-matrices, embody the Galois symmetry as well.
} 
\end{sidewaystable}

Another example is the derived MTC from $\mathsf{M}(3,5)$.
The Hecke images of $\mathsf{M}(3,5)$ include the affine algebras $(A_1,3)$ and $(C_3,1)$. The $\mathsf{M}(3,5)$ MTC has a tensor product structure, which should be maintained under Hecke operations.
Furthermore, the anti-semion in $\mathsf{M}(3,5)$ is a simple current of order 2 and has counterparts in the Hecke images.
Among the fusion rules we focus on two types of fusion, which are referred to as type I and type II. We then compute the $F$-matrices of the correlators.
The field contents and fusion rules are listed in Table \ref{tab:M35}. 
In each theory, the fields in the two types of fusion sit on the (anti-)semion orbit, and the conformal blocks have the same intermediate channels by eq\eqref{fusion_rule:JJ}.
In type I fusion, the $F$-matrices follow from eq\eqref{00,11}:
\beqs
\begin{align}
F^{\mathsf{M}(3,5)}
\begin{bmatrix}
\phi_{\frac{-1}{20}} & \phi_{\frac{-1}{20}} \\ 
\phi_{\frac{-1}{20}} & \phi_{\frac{-1}{20}} 
\end{bmatrix}
&=g
\begin{pmatrix}
1 & * \\ * & -1
\end{pmatrix} ~,\\
F^{\mathsf{\tilde M}(3,5)}
\begin{bmatrix}
\tilde\phi_{\frac{1}{20}} & \tilde\phi_{\frac{1}{20}} \\ 
\tilde\phi_{\frac{1}{20}} & \tilde\phi_{\frac{1}{20}} 
\end{bmatrix}
&=g
\begin{pmatrix}
1 & * \\ * & -1
\end{pmatrix} ~,\\
F^{(A_1,3)}
\begin{bmatrix}
\varphi_{\frac{3}{20}} & \varphi_{\frac{3}{20}} \\ 
\varphi_{\frac{3}{20}} & \varphi_{\frac{3}{20}} 
\end{bmatrix}
&=-\frac{1}{g}
\begin{pmatrix}
1 & * \\ * & -1
\end{pmatrix}~,\\
F^{(C_3,1)}
\begin{bmatrix}
\psi_{\frac{7}{20}} & \psi_{\frac{7}{20}} \\ 
\psi_{\frac{7}{20}} & \psi_{\frac{7}{20}} 
\end{bmatrix}
&=-\frac{1}{g}
\begin{pmatrix}
1 & * \\ * & -1
\end{pmatrix} ~.
\end{align}
\eeqs
In type II fusion, the $F$-matrices are evaluated as
\beqs
\begin{align}
F^{\mathsf{M}(3,5)}
\begin{bmatrix}
\phi_{\frac{1}{5}} & \phi_{\frac{1}{5}} \\ 
\phi_{\frac{1}{5}} & \phi_{\frac{1}{5}} 
\end{bmatrix}
&=-g
\begin{pmatrix}
1 & * \\ * & -1
\end{pmatrix} ~,\\
F^{\mathsf{\tilde M}(3,5)}
\begin{bmatrix}
\tilde\phi_{\frac{4}{5}} & \tilde\phi_{\frac{4}{5}} \\ 
\tilde\phi_{\frac{4}{5}} & \tilde\phi_{\frac{4}{5}} 
\end{bmatrix}
&=-g
\begin{pmatrix}
1 & * \\ * & -1
\end{pmatrix} ~,\\
F^{(A_1,3)}
\begin{bmatrix}
\varphi_{\frac{2}{5}} & \varphi_{\frac{2}{5}} \\ 
\varphi_{\frac{2}{5}} & \varphi_{\frac{2}{5}} 
\end{bmatrix}
&=\frac{1}{g}
\begin{pmatrix}
1 & * \\ * & -1
\end{pmatrix}~,\\
F^{(C_3,1)}
\begin{bmatrix}
\psi_{\frac{3}{5}} & \psi_{\frac{3}{5}} \\ 
\psi_{\frac{3}{5}} & \psi_{\frac{3}{5}} 
\end{bmatrix}
&=\frac{1}{g}
\begin{pmatrix}
1 & * \\ * & -1
\end{pmatrix} ~.
\end{align}
\eeqs
Because of the aforementioned gauge dependence, we do not spell out the off-diagonal entries but denote them by asterisks instead. 
Notice the Frobenius maps between the algebraic numbers $g$ and $-1/g$.
\begin{align}
f_{40,3} (g) = f_{40,7} (g) = -1/g .
\end{align}
For the $\mathsf{M}(3,5)$ theory, we justify that $f_{N,p}$ interpolates the $F$-matrices in the effective picture and the Hecke image under $\mathsf{T}_p$, as claimed.

We are curious how the FSIs are related in the image theories. Later on, the general treatment is based on the picture of effective central charge.
In eq\eqref{FSI_d} the product $\kappa_i \, d_i$ seems an instructive combination, and there is the Galois relation
\begin{align}
\kappa_i^{(p)}\,d^{(p)}_i
= f_{N,\bar p}(\kappa_i \, d_i)
= \kappa_i \, f_{N,\bar p}(d_i)
\end{align}
according to Appendix \ref{appen:results}.
Given the $F$-matrices in the original theory, we acquire their Galois conjugates by making the replacement
\begin{align}
\left\{d_i \right\} \rightarrow \left\{ f_{N, \bar p}( d_i ) \right\}
\end{align}
in all the occurrences of $d_i$ \cite{Freedman}.
The new values obtained are the counterparts in the Hecke image theory under $\mathsf{T}_p$. 

There remains a subtlety about the number field of data in MTC. The solution to the polynomial equations involves $\left\{\sqrt{d_i} \right\}$ under the symmetric normalization. However, $\IQ[\sqrt{d}]$ is a non-abelian extension, which cannot be acted on by the Frobenius map. 
In this case it is not straightforward to find the Galois conjugates of duality matrices.
As pointed out in \cite{Rowell:2009}, all the data of MTC can be presented over certain finite-degree Galois extension of $\mathbb{Q}$, probably over an abelian Galois extension of $\mathbb{Q}$ if normalized appropriately.
That being said, every modular category defined over $\IC$ is conjectured to have a cyclotomic defining number field \cite{Snyder:non_cyclotomic}. The conjecture is restated in \cite{Davidovich:2013}.
If the conjecture holds, one can avoid the non-abelian extension $\IQ[\sqrt{d}]$ and restrict the $F$-matrices in $\IQ[\xi_N]$. The Frobenius maps are then applied unambiguously.

\subsection{Braiding Matrices} 
\label{subsec:Braiding}
The braiding matrices ($B$-matrices) describe unitary transformations of degenerate ground states when the positions of the anyons are fixed.
Akin to the fusion, they exhibit Galois relations upon the Hecke operations. The Frobenius-Schur indicators play a ubiquitous role in such relations.

\begin{figure}[h]
		\centering
 \includegraphics[width=1.6in]{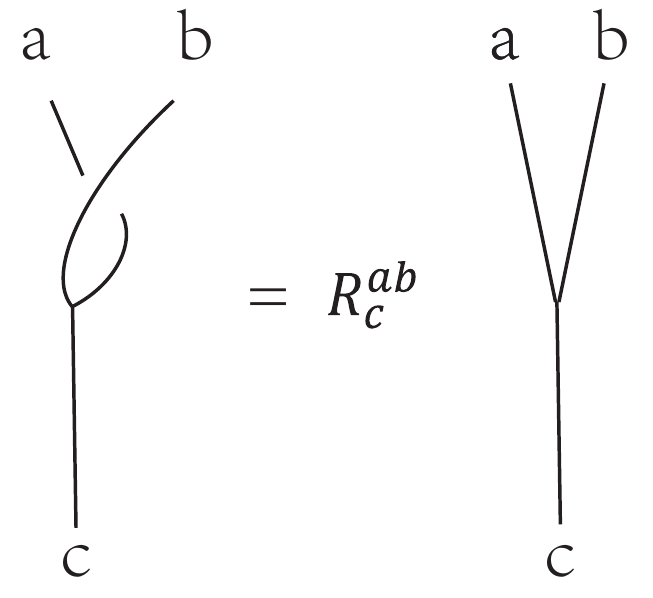}
        \caption{Braiding eigenvalues. }
            \label{fig:R_matrix}
\end{figure}

The braiding matrix is linked to the eigenvalues of R-move by similarity transformation ,as shown in eq\eqref{F_Omega_F}. These eigenvalues arise from interchanging two particles, as illustrated by Figure \ref{fig:R_matrix}. They serve as one-dimensional representations of the braid group, and indicate the statistics of anyons. (The $B$-matrices do not commute and imply non-abelian statistics.)
The braiding eigenvalues are more accessible than the braiding matrices, because they are just square roots of mutual locality factors and are gauge invariant.
Again we restrict ourselves to the fusion rules $| {N_{ab}}^c |\leq 1$. In terms of CVO, the vector space $V_{ab}^c$ is at most one-dimensional for any $a,b,c \in \CI$.

The FSIs occur in the study of braiding matrices, like fusion matrices. When solving the pentagon equations, one cannot fully specify the signs of $6j$ symbols. The signs depend on the FSIs and are chosen correctly by solving the hexagon equations.
For any field $a$, the braiding eigenvalue $R_0^{aa}$ is the phase obtained when two identical particles $a$ are exchanged:
\begin{align}\label{R_aa0}
R_0^{a a} = \kappa_a \,\theta_a^{-1} = \kappa_a \,\me(-h_a).
\end{align}
Here, the FSI $\kappa_a$ is $\pm 1$ if the field $a$ is self-conjugate, and is $0$ if it is complex.
The FSI can be interpreted in terms of angular momentum. For a composite object of zero topological charge formed by two identical anyons, the FSI tells whether its total angular momentum is even or odd, as is evident from eq\eqref{R_aa0} \cite{Kitaev:anyon}. 
Assuming rotational invariance, a rotation of the composite object by $\pi$ is the same as exchanging the two anyons with physical spin $s_a \equiv h_a \pmod{1}$. The rotation then results in a phase factor $e^{\mi \pi s_a}e^{\mi \pi s_a} R_0^{aa}=\kappa_a$ for the whole system. Therefore, $\kappa_a=\pm 1$ determines the parity of the total angular momentum. 
Bantay derives the expression for FSI from the trace of the braiding operator, and finds that 
$\kappa_J = \theta_J^2$ for the simple current $J$ in unitary RCFT \cite{Bantay:FSI}. 
Hence, $\kappa_J =1$ if $h_J$ is an integer or half-integer, while $\kappa_J =-1$ if $ h_J \equiv \pm \frac{1}{4} \pmod{1}$.
To incorporate non-unitary theories, the expression for $\kappa_J$ needs slight modification. By Appendix \ref{appen:results}, the FSI reads
\begin{align}
\kappa_J = \me\big[Q_J(0)\big]\, \theta_J^2.
\end{align}
It generalizes Bantay's formula with a phase factor from the monodromy charge of the vacuum. The FSI of the generic primary field $a$ reads
\begin{align}\begin{split}
\kappa_a &= \sum_{r,s} {}_0{N_{rs}}^a \,
\rho(S)_{0r} \rho(S)_{0s} \, \theta_s^2 \theta_r^{-2} \\
&= \sum_{r,s} {}_0{N_{rs}}^a \,
\rho(S)_{0r} \rho(S)_{0s}  \rho(T)_{ss}^{2} \rho(T)_{rr}^{-2},
\end{split}\end{align}
which solely depends on the modular representation.
Using the modular data, we get the FSI
\begin{align}\label{FSI:eff}
\tilde\kappa_a (\ell^2,J) = \me\big[Q_J(a) + Q_J(0)\big] \, \kappa_a 
\end{align}
in the effective picture $(\ell^2,J)$. As demonstrated before, the map $f_{N,p}$ connects precisely the modular data in the effective picture $(\ell^2,J)$ and the Hecke image theory under $\mathsf{T}_p$. Moreover, it acts on the integer $\tilde\kappa_a$ trivially.
\begin{align}\label{FSI:image}
\kappa_{a}^{(p)}=f_{N,p}\big( \tilde\kappa_a (\ell^2,J) \big) =\tilde\kappa_a (\ell^2,J).
\end{align}
That being said, $\kappa_{a}^{(p)}$ does not depend on specific choice of $p$, as long as ${\bar p}^2 \equiv \ell^2 \pmod{N}$.

We now explore the braiding eigenvalues like $R_b^{aa}$.
In the last section, we computed the four point function $\langle aaaa\rangle$ and studied its fusion matrix, in the case with no more than two channels. 
The braiding symmetry of $\langle aaaa\rangle$ is characterized by the eigenvalues $R_b^{aa}$.
When ${N_{aa}}^b=1$, $R_b^{aa}$ has a compact expression
\begin{align}\label{R_aab}
R_b^{aa} = \theta_a^{-1} \sum_{r,s} {}_0{N_{rs}}^a \,
\rho(S)_{br} \rho(S)_{0s}\, \theta_s^2 \theta_r^{-2} .
\end{align}
This formula holds in general, no matter how many fusion channels there are.
We translate the expression to the effective picture and the Hecke image respectively. It is not difficult to verify that
\begin{align}\label{R_baa:Galois1}
\tilde R_b^{aa} &= f_{N,\ell^2} \big( R_b^{aa} \big) , \\
\label{R_baa:Galois2}
R_b^{aa} {}^{(p)} &= f_{N,p} \big( \tilde R_b^{aa} \big) .
\end{align}

An instructive example is the Hecke image of the Ising model under $\mathsf{T}_p$, which sits in the $\text{spin}(p)$ MTC. With the same notation in Section \ref{subsubsec:Galois_fusion}, we have the twists
\begin{align}
\theta_{\psi}^{(p)} =-1, \qquad 
\theta_{\sigma}^{(p)} =\me\Big(\frac{p}{16}\Big).
\end{align}
The symbol $(p)$ labels the $\text{spin}(p)$ MTC and is omitted for the Ising model itself. The spin field $\sigma^{(p)}$ has the nontrivial FSI 
\begin{align}\label{FSI:spin(p)}
\kappa_{\sigma}^{(p)} =
\Big(\frac{2}{p}\Big)=
\begin{cases} 
1, & \text{if } p \equiv \pm 1\pmod{8},\\
-1, & \text{if } p \equiv \pm 3\pmod{8}.
\end{cases} 
\end{align}
As we know, the FSI for a primary field is $1,0$ or $-1$ if the field is real, complex or quaternionic (a.k.a. pseudo-real) respectively \cite{Rowell:2009}.
The values of $\kappa_{\sigma}^{(p)}$ demonstrate the mathematical fact that the spinor representations of $\text{spin}(p)$ are quaternionic when $p\equiv\pm 3 \pmod{8}$.
For $p\in \big(\IZ/48\IZ)^{\times}$, a little arithmetic verifies that  
\beqs
\begin{alignat}{2}
p^2 &\equiv 1 \pmod{48},  \qquad  &\text{if } p \equiv \pm 1 \pmod{8},\\
p^2 &\equiv 25 \pmod{48},   \qquad &\text{if } p \equiv \pm 3 \pmod{8}.
\end{alignat}
\eeqs
With the Ising fusion rule, there are a total of 16 theories divided into two groups according to the FSI of the spinor.
They correspond to two effective pictures for the Ising model, labeled by $(\ell^2,J)=(1,I)$ and $(\ell^2,J)=(25,\psi)$ respectively.
The monodromy charges under the current $I$ are trivial; while under $J = \psi$ the charges are
\begin{align}
\me\big[Q_{\psi}(I)\big]=\me\big[Q_{\psi}(\psi)\big]=1,\qquad
\me\big[Q_{\psi}(\sigma)\big]=-1.
\end{align}
The FSIs $\kappa_{\sigma}^{(p)}$ are reproduced with these monodromy charges and fall into the two effective pictures.
Following eq\eqref{R_baa:Galois1} and \eqref{R_baa:Galois2}, the nontrivial braiding eigenvalues are
\beqs\label{R:IsingHecke}
\begin{align}
R_0^{\psi\psi} {}^{(p)} &= -1 ~,\\
R_{\sigma}^{\sigma\psi} {}^{(p)} &=R_{\sigma}^{\psi\sigma} {}^{(p)} = \me\Big(-\frac{p}{4}\Big) = -\mi^p ~,\\
R_0^{\sigma\sigma} {}^{(p)} &=
\Big(\frac{2}{p}\Big) \cdot \me\Big(-\frac{p}{16}\Big)~,\\
R_{\psi}^{\sigma\sigma} {}^{(p)} &=
\Big(\frac{2}{p}\Big) \cdot \me\Big(\frac{3p}{16}\Big)~.
\end{align}
\eeqs
A similar result is due to Kitaev \cite{Kitaev:anyon}.

Lastly we turn to $\mathsf{M}(3,5)$, whose Hecke images are computed in Section \ref{subsec:unified_method} as
\begin{align}
\mathsf{T}_3 \chi^{\mathsf{M}(3,5)} &= \chi^{(A_1,3)}, \\
\mathsf{T}_7 \chi^{\mathsf{M}(3,5)} &= \chi^{(C_3,1)}.
\end{align}
Table \ref{tab:M35} lists the field content and the fusion rules in these theories. 
Besides the fusion symmetries, $\mathsf{M}(3,5)$ has the following braiding properties. The primary fields are labeled by $a=(r,s)$ as usual.
\begin{align}\def\arraystretch{1.2}
\begin{array}
{c@{\hspace{0.3cm}}|@{\hspace{0.3cm}}c@{\hspace{0.3cm}}@{\hspace{0.3cm}}c@{\hspace{0.3cm}}@{\hspace{0.3cm}}c@{\hspace{0.3cm}}@{\hspace{0.3cm}}c@{\hspace{0.3cm}}}
\toprule 
\multicolumn{5}{c}{\mathsf{M}(3,5)} \\ \midrule
a & (1,1)  & (2,1) & (3,1) & (4,1) \\ \hline
\theta_a & 1 & \me\big(\frac{-1}{20}\big) & \me\big(\frac{1}{5}\big) & \me\big(\frac{3}{4}\big) \\
\kappa_a & 1 & 1 & 1 & 1 \\ 
R_0^{aa} & 1 & \me\big(\frac{1}{20}\big) & \me\big(\frac{-1}{5}\big) & \me\big(\frac{-3}{4}\big) \\ 
\bottomrule
\end{array} 
\end{align}
We provide the quantities needed to compute the braiding eigenvalues for $(A_1,3)$ and $(C_3,1)$, as well as the effective picture of $\mathsf{M}(3,5)$.
In general let $a$ be any primary field in the original RCFT.
The effective picture $(\ell^2, J)$ amounts to the combined action $a \mapsto \pi_{\ell}Ja$, or equivalently $ J\pi_{\ell}a$ since $J$ and $\pi_{\ell}$ commute.
In $(A_1,3)$ the symbol $j$ means the spin-$j$ representation as usual; 
while in $(C_3,1)$ the primary fields are labeled by the null root $\alpha_0$ and the simple roots $\alpha_1,\alpha_2,\alpha_3$.
\begin{align}\def\arraystretch{1.2}
\begin{array}
{c@{\hspace{0.3cm}}|@{\hspace{0.3cm}}c@{\hspace{0.3cm}}@{\hspace{0.3cm}}c@{\hspace{0.3cm}}@{\hspace{0.3cm}}c@{\hspace{0.3cm}}@{\hspace{0.3cm}}c@{\hspace{0.3cm}}}
\toprule 
\multicolumn{5}{c}{\tilde{\mathsf{M}} (3,5)} \\ \midrule
a & (1,1)  & (2,1) & (3,1) & (4,1) \\ 
\pi_{\ell}Ja & (2,1)  & (1,1) & (4,1) & (3,1) \\ \hline
\tilde\theta_a & 1 & \me\big(\frac{1}{20}\big) & \me\big(\frac{4}{5}\big) & \me\big(\frac{1}{4}\big) \\ 
\tilde\kappa_a & 1 & -1 & 1 & -1 \\ 
\tilde R_0^{aa} & 1 & -\me\big(\frac{-1}{20}\big) & \me\big(\frac{-4}{5}\big) & -\me\big(\frac{-1}{4}\big)  \\ 
\bottomrule
\end{array}
\end{align}
As a consistency check, the twists $\tilde\theta_a$ are also evaluated by shuffling the field content, namely $\tilde h_a \equiv \mathfrak{h}_{\pi_{\ell}Ja} \pmod{1}$.
After bringing $\mathsf{M}(3,5)$ to its effective picture, we then perform the $f_{N,p}$ map to obtain quantities in the Hecke image, such as the twists, FSIs and braiding eigenvalues etc.
\begin{align}\def\arraystretch{1.2}
\begin{array}
{c@{\hspace{0.3cm}}|@{\hspace{0.3cm}}c@{\hspace{0.3cm}}@{\hspace{0.3cm}}c@{\hspace{0.3cm}}@{\hspace{0.3cm}}c@{\hspace{0.3cm}}@{\hspace{0.3cm}}c@{\hspace{0.3cm}}}
\toprule 
\multicolumn{5}{c}{(A_1,3)} \\ \midrule
j & 0  & \frac{1}{2} & 1 & \frac{3}{2} \\ \hline
\theta_j & 1 & \me\big(\frac{3}{20}\big) & \me\big(\frac{2}{5}\big) &  \me\big(\frac{3}{4}\big)\\ 
\kappa_j & 1 & -1 & 1 & -1 \\ 
R_0^{jj} & 1 & -\me\big(\frac{-3}{20}\big) & \me\big(\frac{-2}{5}\big) & -\me\big(\frac{-3}{4}\big)  \\ 
\bottomrule
\end{array}
\end{align}

\begin{align}\def\arraystretch{1.2}
\begin{array}
{c@{\hspace{0.3cm}}|@{\hspace{0.3cm}}c@{\hspace{0.3cm}}@{\hspace{0.3cm}}c@{\hspace{0.3cm}}@{\hspace{0.3cm}}c@{\hspace{0.3cm}}@{\hspace{0.3cm}}c@{\hspace{0.3cm}}}
\toprule 
\multicolumn{5}{c}{(C_3,1)} \\ \midrule
\alpha & \alpha_0  & \alpha_1 & \alpha_2 & \alpha_3 \\\hline
\theta_{\alpha} & 1 & \me\big(\frac{7}{20}\big) & \me\big(\frac{3}{5}\big) & \me\big(\frac{3}{4}\big) \\ 
\kappa_a & 1 & -1 & 1 & -1 \\ 
R_0^{\alpha\alpha} & 1 & -\me\big(\frac{-7}{20}\big) & \me\big(\frac{-3}{5}\big) & -\me\big(\frac{-3}{4}\big)  \\ 
\bottomrule
\end{array}
\end{align}

So far we have seen the Galois relations in $R^{aa}_b$.
Without the expression for $R^{ab}_c$, it seems difficult to find the Galois conjugates of general braiding eigenvalues, though the same Galois symmetry is expected to hold.
Nevertheless, $R^{ab}_c$ squares to the mutual locality factor.
\begin{align}
\big( R^{ab}_c \big)^2 = \me (h_c - h_a - h_b) ,
\qquad a,b,c \in \CI ~.
\end{align}
For $R^{ab\,(p)}_c$ in the image theory under $\mathsf{T}_p$, the above relation implies
\begin{align}\label{eff_to_image}
\Big( R^{ab\,(p)}_c \Big)^2 
= f_{N,p} \left( \big( \tilde R^{ab}_c \big)^2 \right) ,
\end{align}
due to the Galois relation between the twists.
In a similar vein, we establish
\begin{align}\label{preimage_to_eff}
\big( \tilde R^{ab}_c \big)^2 =  f_{N,\ell^2} \left( \big( R^{ab}_c \big)^2 \right) 
\end{align}
for $ \tilde R^{ab}_c $ in the effective picture, where the selection rule eq\eqref{selection_rule} is used.
They support the conjecture that the braiding eigenvalues are related by the same Frobenius map for the modular representations.
We have the neat relations
\begin{align}
 \tilde R^{ab}_c  &=  f_{N,\ell^2} \left( R^{ab}_c \right) , \\
 R_{c}^{ab} {}^{(p)} &=  f_{N,p} \left(  \tilde R^{ab}_c \right) .
\end{align}
They are argued as follows. 
Because of the identical fusion rules, the braiding eigenvalues in the effective picture saturate the same hexagon equations, but with Galois-conjugate $F$-matrices inserted. The solved braiding eigenvalues are then related by the same Galois symmetry for the $F$-matrices. To be precise, 
$$\left\{f_{N,\ell^2}\big(R^{ab}_c\big) \, \big| \, a,b,c \in \CI \right\}$$
constitute the solution in the effective picture. Similarly, 
$$\left\{f_{N,\bar p}\big(R^{ab}_c\big) \, \big| \, a,b,c \in \CI \right\}$$
are the braiding eigenvalues for the Hecke image under $\mathsf{T}_p$.
As compositions of the $F$-matrices and the braiding eigenvalues, the $B$-matrices obey the same Frobenius map.

Based on the study of fusion and braiding, we finally reach the conclusion that the Galois symmetry in the Hecke relations also connects the duality quantities of RCFTs.

\section{Conclusions and Outlook}
\label{sec:Conclusions}

The Hecke operators reveal novel relations between the characters of a number of interesting RCFTs with small numbers of independent characters.  In addition to relating characters, the Hecke operators also induce Galois symmetries between modular representations, thus connecting analytic and algebraic number theory in the context of RCFT.  It is natural to wonder whether these Hecke relations
are a sign of a deeper number theoretic relation between certain RCFTs. A preliminary step towards answering this question is to ask whether the MTCs of two RCFTs whose characters are related by Hecke relations are isomorphic.
In this paper we have shown that this is the case for special classes of MTCs related to minimal models or with only two fusion channels by utilizing the duality properties in the Hecke image theory where the Galois symmetry relating modular representations is extended to the duality matrices.
In our framework, the picture of effective central charge occurs as a significant intermediate step, and is useful for identifying unitary Hecke images. Specifically, physical quantities in the effective picture and the initial theory are related through Galois inner automorphism and simple current permutation. 
For the Hecke image under $\mathsf{T}_p$, modularity and the duality properties are then deduced by acting with the Frobenius map $f_{N,p}$ on the data in the effective picture. As part of this procedure we also provided a unified study of the Frobenius-Schur indicator of the MTC and its Hecke image. 
The equations of $T$, $S$, $F$, $B$ transformations show that all the conformal blocks on every genus form a representation of the whole duality groupoid \cite{Moore:polynomial}. The modular group is a subgroup of the duality groupoid. It makes sense that the duality matrices obey the same Galois symmetry as for the generators of the modular group. It would be interesting to pursue how the duality groupoid encodes the Galois symmetry induced by Hecke relations in future work. We also formulated a relation between the modular representations of the minimal models $\mathsf{M}(2,k+2)$ and the simple-current reduced affine algebras $(A_1,k)_{\frac{1}{2}}$, connecting non-unitary RCFTs to unitary ones by Galois symmetry. This relation could prove useful in condensed matter theory  where the unitarity of the theory is determined by tuning the couplings in the Hamiltonian.

There are further applications of Hecke operators in RCFT that are interesting to explore. In \cite{Hecke:2018} it was pointed out that the characters of the $c=47/2$ VOA with Baby Monster symmetry are Hecke images of Ising model
characters. This Hecke relation has recently been extended to a relation between the characters of other minimal models and parafermion theories and characters of VOAs with other sporadic automorphism groups \cite{bhllr}. It would also be interesting to investigate the action of Hecke operators on the characters of intermediate vertex algebras and their relation to RCFT characters. The central open problem is to understand from a more fundamental point of view why
Hecke relations exist between RCFT characters and whether this is a signal of some new number theoretical aspects of RCFT which are still to be understood.

\section*{Acknowledgements}
We would like to thank S. Gukov, J. Heckman, C. Kane, M. Levin, G. Moore, S. Mukhi, E. Rowell, N. Snyder and Z. Wang for helpful discussions and correspondence.
We thank C. Ferko for reading the manuscript and offering advice.
J.H. and Y.W. acknowledge support from the NSF \footnote{Any opinions, findings, and conclusions or recommendations expressed in this material are those of the author(s) and do not necessarily reflect the views of the National Science Foundation.} under grant PHY 1520748.

\appendix

\section{Jacobi Symbol}
\label{appen:Jacobi_symbol}
The Jacobi symbol appears frequently in the Frobenius map on abelian extensions of $\IQ$. 
In the discussion of MTC, the Frobenius map takes the fusion matrix to its counterpart in a Galois-conjugate MTC.
Here we provide a general treatment for the Jacobi symbol.

First define the Legendre symbol as a special case.
The Legendre symbol $\big(\frac{a}{p}\big)$ is defined for all integers $a$ and all odd primes $p$ by
\begin{align}
\left(\frac{a}{p}\right) := 
\begin{cases} 
0 & \text{if } a \equiv 0 \pmod{p},\\
1 & \text{if } a \not\equiv 0\pmod{p} \text{ and for some integer } x\colon\;a\equiv x^2\pmod{p},\\
-1 & \text{if } a \not\equiv 0\pmod{p} \text{ and there is no such } x.
\end{cases} 
\end{align}

The Jacobi symbol is a generalization of the Legendre symbol. For any integer $ a$ and any positive odd integer $n$, let $n=p_1^{r_1}\cdots p_m^{r_m}$ be the prime factorization.
The Jacobi symbol $\big( \frac{a}{n} \big)$ is defined as the product of the Legendre symbols corresponding to the prime factors of $n$.
\begin{align}
\Big(\frac{a}{n}\Big) :=
\Big(\frac{a}{p_1}\Big)^{r_1} \cdots
\Big(\frac{a}{p_m}\Big)^{r_m} ~.
\end{align}
Following the definition of the Jacobi symbol, we have the obvious properties:
\begin{align}
\Big(\frac{a+bn}{n}\Big)&=\Big(\frac{a}{n}\Big)~, \label{JacobiSymbol:1}\\
\Big(\frac{a}{n}\Big)\Big(\frac{b}{n}\Big)&=\Big(\frac{ab}{n}\Big)~,\label{JacobiSymbol:2}\\
\Big(\frac{a}{m}\Big)\Big(\frac{a}{n}\Big)&=\Big(\frac{a}{mn}\Big)~\label{JacobiSymbol:3}.
\end{align}
The Jacobi symbol obeys the profound law of quadratic reciprocity: if $m$ and $n$ are odd positive coprime integers, then
\begin{align}\label{quadratic_reciprocity}
\Big(\frac{n}{m}\Big)\Big(\frac{m}{n}\Big)
= (-1)^{\frac{m-1}{2} \cdot \frac{n-1}{2}}~.
\end{align}

The main result we need in the analysis in the text is
\begin{align}\label{proposition:Jacobi}
f_{L,\ell} \big( \sqrt{K} \big) =  \Big(\frac{K}{\ell}\Big)\sqrt{K} ~,
\qquad \ell \in (\mathbb{Z}/L\mathbb{Z})^{\times},
\end{align}
where any prime factor of $K$ also divides $L$. 
This formula connects the Jacobi symbol to the Frobenius map on quadratic irrational numbers.

In most cases, $L$ is even and hence $\ell$ must be odd. We can extend $\IQ[\xi_L]$ to a larger cyclotomic field $\IQ[\xi_{2^m L}]$ which does not affect $f_{L,\ell}$. Without loss of generality, we assume $L \equiv 0\pmod{8}$.
To prove eq\eqref{proposition:Jacobi}, we first show that for any prime factor $p$ of $K$ there is
\begin{align}\label{Frobenius:sqrt}
f_{L,\ell}( \sqrt{p} )=  \Big(\frac{p}{\ell}\Big)\sqrt{p} 
\end{align}
with odd $\ell$ \cite{Coste:1993af}.
There are two cases depending on the parity of $p$.
When $p$ is an odd prime number, we exploit the Gauss sum
\begin{align}
\mathfrak{G}(b;p) := \sum_{i=0}^{p-1} \xi_p^{i^2 b},
\qquad b \in \IZ.
\end{align}
Since any prime factor of $K$ also divides $L$ as assumed, $\mathfrak{G}(b;p)$ takes values in $\IQ[\xi_L]$. 
The Gauss sum is related to the quadratic integer $\sqrt{p}$ via
\begin{align}\label{sqrt_p}
\sqrt{p} = \omega \, \mathfrak{G}(1;p) ~,
\end{align}
where
\begin{align}
\omega = 
\begin{cases} 
1 & \text{if } p \equiv 1\pmod{4},\\
-\mi & \text{if } p \equiv -1\pmod{4}.
\end{cases} 
\end{align}
It can be shown that $f_{L,\ell}( \omega )=-\omega$ only when
$\ell \equiv p  \equiv -1 \pmod{4}$,
otherwise $f_{L,\ell}( \omega )=\omega$. This result is rephrased as
\begin{align}
f_{L,\ell}( \omega )  =  (-1)^{\frac{\ell-1}{2} \cdot \frac{p-1}{2}} \omega ~.
\end{align}
Recall a property of the Gauss sum
\begin{align}
f_{L,\ell}\big( \mathfrak{G}(1;p) \big)
\equiv \mathfrak{G}(\ell;p) 
= \Big(\frac{\ell}{p}\Big) \mathfrak{G}(1;p),
\end{align}
where $p|L$. Acting $f_{L,\ell}$ on $\sqrt{p}$, we find
\begin{align}\begin{split}
\frac{ f_{L,\ell}( \sqrt{p} ) }{\sqrt{p}}
&= \frac{ f_{L,\ell}( \omega ) }{\omega} \cdot 
\frac{ f_{L,\ell}\big( \mathfrak{G}(1;p) \big) }{\mathfrak{G}(1;p)}
=\frac{ f_{L,\ell}( \omega ) }{\omega} \cdot 
\frac{ \mathfrak{G}(\ell;p) }{\mathfrak{G}(1;p)} \\
&=  (-1)^{\frac{\ell-1}{2} \cdot \frac{p-1}{2}} \cdot
 \Big(\frac{\ell}{p}\Big) 
 =  \Big(\frac{p}{\ell}\Big)~,
\end{split}\end{align}
where in the last equality we use the law of quadratic reciprocity.
While for $p=2$, notice that
\begin{align}
 \me\Big(\frac{\nu}{8}\Big) + \me\Big(-\frac{\nu}{8}\Big)
 =(-1)^{ \frac{\nu^2-1}{8} }
\sqrt{2}~.
\end{align}
Comparing the results for $\nu=1$ and $\nu=\ell$, we get
\begin{align}
f_{L,\ell}( \sqrt{2} ) 
= (-1)^{ \frac{\ell^2-1}{8} } \sqrt{2} 
=\Big( \frac{2}{\ell} \Big) \sqrt{2} .
\end{align}
Hence, eq\eqref{Frobenius:sqrt} holds as claimed.

There are some RCFTs with odd conductors. We have to take into account odd $L$ when proving the proposition. In this case only $f_{L,\ell}$ with odd $\ell$ needs to be considered.
(If $\ell$ is even, we instead consider the map $f_{L,\ell - L}$, which is equivalent to $f_{L,\ell}$ when acting on $\IQ[\xi_L]$.)
Since $\ell$ is odd, we can again enlarge $\IQ[\xi_L]$ to $\IQ[2^m \xi_L]$ without affecting $f_{L,\ell}$.
The situation then reduces to the case of even $L$, which has been analyzed previously.

Returning to eq\eqref{proposition:Jacobi} we note that
any even power can be taken outside the square root, so we only need to consider $K$ with the prime factorization $K=\prod_j k_j$ where $k_j$ are distinct prime numbers. With the multiplication rule eq\eqref{JacobiSymbol:2}, one has
\begin{align}
 \Big(\frac{K}{\ell}\Big) = \prod_j  \Big(\frac{k_j}{\ell}\Big) 
 =\prod_j  \frac{f_{L,\ell} (\sqrt{k_j}) }{\sqrt{k_j}}
 = \frac{f_{L,\ell} (\sqrt{K}) }{\sqrt{K}}~,
\end{align}
proving the assertion.

\section{Galois Symmetry Interpolated Modular Representations}
\label{appen:Galois}
In this section we explore the Galois connection between the $\mathsf{M}(2,k+2)$ and the $(A_1,k)_{\frac{1}{2}}$ modular representations with odd $k$.
In addition to the conductor $N$, we set $N_0$ to be the least common denominator of the conformal weights. Proposition 5 in \cite{Bantay:2001ni} states that
\begin{align}
N=\mathfrak{e}N_0,
\end{align}
where the integer $\mathfrak{e}$ divides 12. In addition, $\text{gcd}(\mathfrak{e},N_0)=1 \text{ or }2$.

For the basis $\phi_{(u,1)}$ in $\mathsf{M}(2,k+2)$, the representation is determined by 
\beqs
\begin{align}
\rho^{\mathsf{M}}(T)_{u,v} &= \delta_{u,v}\,
\me\left(\frac{(k+2-2u)^2}{8(k+2)} - \frac{1}{24}\right), \\
\rho^{\mathsf{M}}(S)_{u,v} &= \frac{2}{\sqrt{k+2}} (-1)^{1+u+v} 
\sin\Big(\frac{uv}{k+2}2\pi\Big)
\sin\Big(\frac{k+2}{2}\pi\Big).
\end{align}
\eeqs
While the modular representation of $(A_1,k)_{\frac{1}{2}}$ is
\beqs
\begin{align}
\rho^{A}(T)_{l,l'} &= \delta_{l,l'}\,
\me\left(\frac{l^2}{4(k+2)} -\frac{1}{8} \pm \frac{1}{24}\right), \\
\rho^{A}(S)_{l,l'} &= \frac{2}{\sqrt{k+2}} 
\sin\Big(\frac{l l'}{k+2}\pi\Big),
\end{align}
\eeqs
where the signs $\pm$ correspond to the cases $k\equiv \pm 1 \pmod{4}$ respectively and $l,l'$ are odd integers that satisfy $1\leq l,l'\leq k $.

We will not directly apply the Frobenius map $f_{N,-k}$ on $\rho(\gamma)$, for fear that $k$ may not be coprime to the conductor.
Instead, we exploit the techniques in MTC \cite{Rowell:2009}. Rather than working with the CFT-normalized $T$ matrix, we use an appropriate surjective restriction: $(\IZ / N\IZ)^{\times} \rightarrow (\IZ / N_0\IZ)^{\times}$.
Define the diagonal matrix
\begin{align}
\varrho(T) = \me\Big(\frac{c}{24}\Big) \rho(T)
\end{align}
for the $T$ transformation in MTC.
The diagonal entries of $\varrho(T)$ consist of all the twists and take values in $\IQ[\xi_{N_0}]$.
We then have
\begin{align}
\varrho^{\mathsf{M}}(T)_{u,v} &= \delta_{u,v}\,
\me\left(\frac{(k+2-2u)^2-k^2}{8(k+2)} \right), \\
\varrho^A(T)_{l,l'} &=\delta_{l,l'}\,
\me\left( \frac{l^2-1}{4(k+2)} \right).
\end{align}
Since the indices are odd, we find $N_0=k+2$ in both MTCs.
Similarly, we write $\varrho(S)=\rho(S)$. In some sense, the pair $\big( \varrho(T),\varrho(S) \big)$ also characterizes $SL(2,\IZ)$ as $\big( \rho(T),\rho(S) \big)$ does.
The constraints for $\big( \varrho(T),\varrho(S) \big)$ are instead
\beqs
\begin{align}
\varrho(S)^2 &= \CC, \\
\big( \varrho(T)\varrho(S) \big)^3 &=\CC\, \me\Big(\frac{c}{8}\Big) ,
\end{align}
\eeqs
where $\CC$ is the charge conjugation matrix.
We also learn that $\varrho(S)$ is in $\IQ[\xi_{4(k+2) }]$ for both MTCs \cite{Bantay:2001ni,Francesco:2012}. 
Therefore the extension of $\IQ$ by either $\varrho^{\mathsf{M}}(\gamma)$ or $\varrho^A(\gamma)$ leads to $\IQ[\xi_{4(k+2) }]$, on which there exists the Frobenius map $f_{4(k+2),-k}$.

We proceed to act the Frobenius map $f_{4(k+2),-k}$ on $\varrho^A(T)$ and $\varrho^A(S)$ respectively. 
\begin{align}\begin{split}
 f_{4(k+2),-k} \big( \varrho^A(T)_{l,l'} \big) &= \delta_{l,l'}\,
\me\left(- k \frac{l^2-1}{4(k+2)} \right) \\
&= \delta_{l,l'}\,
\me\left(\frac{(k+2-2l)^2-k^2}{8(k+2)} -\Big(\frac{l-1}{2}\Big)^2 \right) \\
&= \delta_{l,l'}\,
\me\left(\frac{(k+2-2l)^2-k^2}{8(k+2)} \right) 
\equiv  \varrho^{\mathsf{M}}(T)_{l,l'} ~.
\end{split}\end{align}
\begin{align}
\begin{split}
 f_{4(k+2),-k} \big( \varrho^A(S)_{l,l'} \big) 
&= \Big(\frac{k+2}{k}\Big) \frac{2}{\sqrt{k+2}} 
\, \mi^{k+1} \sin\Big(-\frac{k l l'}{k+2}\pi\Big) \\
&=  \Big(\frac{2}{k}\Big) \frac{2}{\sqrt{k+2}} 
\, \mi^{k+1} (-1)^{l l'} \sin\Big(l l'\pi-\frac{k l l'}{k+2}\pi\Big) \\
&=  \Big(\frac{2}{k}\Big) \frac{2}{\sqrt{k+2}} 
\, \mi^{k-1} \sin\Big(\frac{l l'}{k+2} 2\pi\Big) 
\equiv \Big(\frac{2}{k}\Big) \varrho^{\mathsf{M}}(S)_{l,l'} ~.
\end{split}
\end{align}
In the derivations we bear in mind that $l,l'$ are odd.
The fusion rule isomorphism is validated by the shown Galois symmetry between the modular $S$ matrices.

From the MTC point of view, $\mathsf{M}(2,k+2)$ is same as the Galois conjugate of $ (A_1,k)_{\frac{1}{2}} $, up to a one-dimensional modular representation of central charge
\begin{align}\begin{split}
c_{\rm tot } &=
c \big[ \mathsf{M}(2,k+2) \big] + k \cdot c \left[ (A_1,k)_{\frac{1}{2}} \right] \\
&= 1-\frac{6k^2}{2(k+2)}+k\Big(\frac{3k}{k+2} \mp 1\Big)=1\mp k.
\end{split}\end{align}
This central charge leads to
\beqs
\begin{align}
\rho^{\text{1d}}(T) &= \me\Big(-\frac{ c_{\rm tot } }{24}\Big)
=  \me\Big(\frac{ \pm k -1 }{24}\Big) 
=\begin{cases} 
\me\big(\frac{ t }{6}\big), & \text{if } k =4t+1,\\
\me\big(-\frac{ t }{6}\big), & \text{if } k =4t-1,
\end{cases} \\
\rho^{\text{1d}}(S) &= \Big(\frac{2}{k}\Big) 
=  (-1)^{ \frac{k^2-1}{8} } = (-1)^t, 
\qquad \text{for } k=4t \pm 1,
\end{align}
\eeqs
where $t$ is a positive integer.
$\rho^{\text{1d}}$ agrees with the one-dimensional representations classified by Lemma 5 of \cite{eholzerII}.

There is another approach to understand this Galois symmetry.
With the effective description, the $(-k)$-th Galois conjugate of $\tilde{\mathsf{M}} (2,k+2)$ differs from $ \rho^{(A_1,k)_{\frac{1}{2}}}$ by a one-dimensional representation, which has central charge
\begin{align}\begin{split}
c'_{\rm tot } &= k \cdot c \big[ \tilde{\mathsf{M}} (2,k+2) \big] + c \left[ (A_1,k)_{\frac{1}{2}} \right] \\
&= k \Big(1 - \frac{3}{k+2} \Big) + \frac{3k}{k+2} \mp 1= k \mp 1.
\end{split}\end{align}

Gannon provides the unitarization of $\mathsf{M}(2,k+2)$, which is slightly different from the above \cite{Gannon:nonunitary}.
Our analysis demonstrates that the unitarization of $\mathsf{M}(2,k+2)$ leads to the $ (A_1,k)_{\frac{1}{2}} $ MTC.


\section{Derivation of MTC Data}
\label{appen:results}
In this appendix we provide the derivation for some data in the Hecke image theory.
We restrict ourselves to RCFTs without degenerate twists.

Denote by $d_b$, $\tilde d_b$ and $d^{(p)}_b$ respectively the quantum dimensions in the original theory, the effective description $(\ell^2, J)$ and the Hecke image under $\mathsf{T}_p$. As always, $\ell$ is odd and ${\bar p}^2 \equiv \ell^2 \pmod{N}$.
For each individual $p$, $\mathsf{T}_p$ produces positive quantum dimensions in the image theory, which offer evidence for unitary RCFTs. 
By definition, $d^{(p)}_b$ takes the form
\begin{align}
d^{(p)}_b = \frac{\rho^{(p)}(S)_{0b}}{\rho^{(p)}(S)_{00}} \, .
\end{align}
It is simplified as follows.
\begin{align}\begin{split}\label{qdim:p}
d^{(p)}_b &= f_{N,p}\left(\frac{\tilde\rho(S)_{0b}}{\tilde\rho(S)_{00}}\right) \\
&=f_{N,p}\circ f_{N,\ell^2} \left(\frac{\rho(S)_{J0,Jb}}{\rho(S)_{J0,J0}}\right) 
= f_{N,\ell} \left(\frac{\rho(S)_{J0,Jb}}{\rho(S)_{J0,J0}}\right) \\
&=\frac{\rho(S)_{\pi_{\ell}J0,Jb}}{\rho(S)_{\pi_{\ell}J0,J0}}
=\frac{\rho(S)_{\mathit{o},Jb}}{\rho(S)_{\mathit{o},J0}}
=\frac{\rho(S)_{\mathit{o},b}}{\rho(S)_{\mathit{o},0}}.
\end{split}\end{align}
In the second line above, we use the fact that $f_{p\ell}$ yields the identity permutation of the fields when $(p\ell)^2 \equiv 1 \pmod{N}$, c.f. Section \ref{subsec:Galois_Permu}. In the last line, the minimal primary is reached from the vacuum by $\mathit{o}=\pi_{\ell}J0$.
Gannon defines $\rho(S)_{\mathit{o},b}/\rho(S)_{\mathit{o},0}$ as the quantum dimension for non-unitary cases, and shows that
\begin{align}
\rho(S)_{\mathit{o},b}/\rho(S)_{\mathit{o},0}\geq 1
\end{align}
\cite{Gannon:nonunitary}. Nevertheless, we stick to the definition eq\eqref{def:q_dim} for quantum dimensions.
Independent of specific choices of $p$, the quantum dimensions $d^{(p)}_b$ are inherently encoded in the initial modular $S$ matrix. The values $d^{(p)}_b \geq 1$ suggest that the image theory is unitary.
While in the effective picture the quantum dimensions are
\begin{align}
\tilde d_b = \frac{\tilde\rho(S)_{0b}}{\tilde\rho(S)_{00}}
= f_{N,\ell^2} \left(\frac{\rho(S)_{J0,Jb}}{\rho(S)_{J0,J0}}\right) 
= \me\big[Q_J(b)-Q_J(0)\big] \,f_{N,\ell^2} \left(\frac{\rho(S)_{0,b}}{\rho(S)_{0,0}}\right) ,
\end{align}
which need not be positive.

In the effective description $(\ell^2,J)$, the FSI reads
\begin{align}
\tilde\kappa_a (\ell^2,J) = \sum_{r,s} {}_0{\tilde N_{rs}}^{~~~a} \, \tilde\rho(S)_{0r} \tilde\rho(S)_{0s} \, \tilde\theta_s^2 \tilde\theta_r^{-2} \, ,
\end{align}
where ${}_0{\tilde N_{rs}}^{~~~a}$ equals ${}_0{N_{rs}}^a $ by eq\eqref{FusionRules:same1}.
We act on this formula by $f_{N,\bar\ell^2}$ and notice that ${}_0{N_{rs}}^a = {}_0{N_{Jr,Js}}^a $ for $J^2 = I$.
\begin{align}\begin{split}
\label{FSI:derivation}
f_{N,\bar\ell^2}\big( \tilde\kappa_a (\ell^2,J) \big) 
&= \sum_{r,s} {}_0{N_{rs}}^a \, \rho(S)_{J0,Jr} \rho(S)_{J0,Js} \, \big(\theta_{Js}\theta_J^{-1}\big)^2 \big(\theta_{Jr}\theta_J^{-1}\big)^{-2} \\
&= \sum_{r,s} {}_0{N_{Jr,Js}}^a \, \me\big[Q_J(Jr)+Q_J(Js)\big] \, \rho(S)_{0,Jr} \rho(S)_{0,Js} \, \theta_{Js}^2 \theta_{Jr}^{-2} \\
&= \me\big[Q_J(a)+Q_J(0)\big] \sum_{r,s} {}_0{N_{Jr,Js}}^a \,\rho(S)_{0,Jr} \rho(S)_{0,Js} \, \theta_{Js}^2 \theta_{Jr}^{-2} \\
& \equiv \me\big[Q_J(a)+Q_J(0)\big] \, \kappa_a 
\end{split}\end{align}
It confirms that $f_{N,\bar\ell^2}\big( \tilde\kappa_a (\ell^2,J) \big) $ and therefore $\tilde\kappa_a (\ell^2,J)$ are integers. The FSIs in the effective description and the original theory are related by
\begin{align} \label{kappa:eff}
\tilde\kappa_a (\ell^2,J) = \me\big[Q_J(a) + Q_J(0)\big] \, \kappa_a \, .
\end{align}
Since the monodromy charges vanish under the trivial simple current, the FSIs are preserved in this particular scenario.
Moreover, the FSI in the image theory is translated invariantly from the effective description.
\begin{align} \label{kappa:p}
\kappa_{a}^{(p)}=f_{N,p}\big( \tilde\kappa_a (\ell^2,J) \big) =\tilde\kappa_a (\ell^2,J) \, .
\end{align}

It remains to study the orbit of the simple current $J$. We assume $J^2=I$, in which case $J$ could play a role in the effective description. For the quantum dimensions, we have
\begin{align}
d_{Jb} = \frac{\rho(S)_{0,Jb}}{\rho(S)_{0,0}}
= \me\big[Q_J(0)\big] \frac{\rho(S)_{0,b}}{\rho(S)_{0,0}} = \me\big[Q_J(0)\big] d_b ~.
\end{align}
On a simple current orbit, the fields have quantum dimensions of the same magnitude. In particular they are equal for unitary RCFTs, in which the vacuum has trivial monodromy charge.
Given the fusion rule $\phi_r \times \phi_s = \sum_b {}_0{N_{rs}}^b \, \phi_b$, we find
\begin{align}
\phi_{Jr} \times \phi_s = \sum_b {}_0{N_{rs}}^b \, \phi_{Jb} \, ,
\end{align}
where ${}_0{N_{Jr,s}}^{Jb} = {}_0{N_{rs}}^b$ by straightforward computation. 
For the FSI, there is
\begin{align}\begin{split}
\kappa_{Jb} &= \sum_{r,s} {}_0{N_{rs}}^{Jb} \, \rho(S)_{0r} \rho(S)_{0s} \, \theta_s^2 \theta_r^{-2} \\
&= \sum_{r,s} {}_0{N_{Jr,s}}^{Jb} \, \rho(S)_{0,Jr} \rho(S)_{0s} \, \theta_s^2 \theta_{Jr}^{-2} \\
&= \me\big[Q_J(0)\big] \sum_{r,s} {}_0{N_{rs}}^b \,\rho(S)_{0r} \rho(S)_{0s} \, \theta_s^2 \theta_r^{-2} \theta_J^2 \\
&= \me\big[Q_J(0)\big] \, \theta_J^2 \, \kappa_b.
\end{split}\end{align}
To see the application of this formula, we return to any individual theory in the $\mathsf{M}(3,5)$ series. In type I and type II fusions, the external fields sit on the (anti-)semion orbit, implying 
\begin{equation}
\theta_J^2 = \me(2 h_J) = -1. 
\end{equation}
With the quantum dimensions and the FSIs inserted, eq\eqref{FSI_d} explains why the overall signs of the $F$-matrices are different for the two types of fusion.

\end{document}